\documentclass[pra,notitlepage,english,superscriptaddress,twocolumn,floatfix]{revtex4-1}
\usepackage{graphicx,epsfig,epstopdf,subfigure,hyperref}
\usepackage{bm}
\pdfoutput=1
\usepackage{amstext}
\usepackage{amssymb}
\usepackage{amsmath}
\usepackage{braket}
\usepackage{color}
\usepackage{babel}

\def\({\left(}
\def\){\right)}
\def\min{\textrm{min}}
\def\max{\textrm{max}}
\def\Tr{\textrm{Tr}}
\expandafter\let\csname eqref\endcsname\relax
\expandafter\let\csname eqref\endcsname\relax
\graphicspath{figures/}
\begin{document}

\title{Splitting a critical spin chain}

\author{Alejandro Zamora}
\affiliation{ICFO - The Institute of Photonic Sciences, Av. C.F. Gauss 3, E-08860 Castelldefels (Barcelona), Spain} 

\author{Javier Rodr\'{\i}guez-Laguna}
\affiliation{ICFO - The Institute of Photonic Sciences, Av. C.F. Gauss 3, E-08860 Castelldefels (Barcelona), Spain}
\affiliation{GISC \& Mathematics Dept., Universidad Carlos III de Madrid, Legan\'es, Spain}

\author{Maciej Lewenstein}
\affiliation{ICFO - The Institute of Photonic Sciences, Av. C.F. Gauss 3, E-08860 Castelldefels (Barcelona), Spain}
\affiliation{ICREA-Instituci\'{o} Catalana de Recerca i Estudis
Avan\c{c}ats, 08010 Barcelona, Spain}

\author{Luca Tagliacozzo}
\affiliation{ICFO - The Institute of Photonic Sciences, Av. C.F. Gauss 3, E-08860 Castelldefels (Barcelona), Spain}

\begin{abstract}
We study a quench protocol that conserves the entanglement spectrum of a bipartition of a quantum system. As an example we consider the splitting of a critical Ising chain in two chains, and compare it with the well known case of joining of two chains. We show that both the out of equilibrium time evolution of global properties and the equilibrium regime after the quench of local properties are different in the two scenarios. Since the two quenches only differ in the presence/absence of the conservation of the entanglement spectrum, our results suggest that this conservation plays a fundamental role in both the out-of-equilibrium dynamics and the subsequent equilibration mechanism. We discuss the relevance of our results to the next generation of quantum simulators.
\end{abstract}

\maketitle


\section{Introduction}

The dynamical evolution of an isolated quantum system is governed by a
unitary operator and is, consequently, reversible. Therefore, one
might think that irreversibility and thermalization should only appear
through the system-environment interaction \cite{von_neumann.29}. 
For a small region inside a large isolated quantum system, a legitimate environment is the system itself. In particular,
it is important to understand,
under which conditions the large-time out of equilibrium evolution of an
isolated system will lead to a thermal state of the small region. Although the decoherence time of most experimental systems
is too short for an effective study of that regime, recent advances in
cold atomic physics \cite{lewenstein.book} have allowed to experimentally address such situations and have boosted renewed interest in the theoretical understanding of these phenomena
\cite{kinoshita_quantum_2006,cheneau_light-cone-like_2012,gring_relaxation_2012,trotzky_probing_2012,langen_local_2013}. The
experiments have been complemented with theoretical insights
\cite{deutsch.91,sredniki.94,rigol.08,polkovnikov.11}, which have
brought about interesting ramifications of the problem, ranging from
 quantum information and entanglement to the issue of integrability in quantum systems.

 In the context of the low energy physics of a many body quantum systems, entanglement has emerged as a privileged tool to characterize quantum phases. In 1D, for example, the scaling of entanglement allows to distinguish between gapped systems and critical systems, and the structure of the entanglement spectrum allows to identify symmetry protected topological phases.
Here we try to analyze the effects of the conservation of the entanglement in the out-of-equilibrium evolution after a quantum quench. 

Conserved quantities play a very special role in Physics.
 In classical mechanics, they allow to define  integrable systems as those systems that possess as many conserved quantities 
as degrees of freedom. In quantum mechanics, this concept is hard to generalize. The expectation value of any operator that commutes with the system Hamiltonian is conserved. In particular, arbitrary powers of the Hamiltonian itself (that in general can define independent operators)  are  conserved during the out-of equilibrium dynamics.
This means that  a generic quantum system  possesses as many conserved quantities as degrees of freedom, and we still miss a proper definition of integrable quantum systems.

Furthermore, when considering \emph{local equilibration},  the equilibration of a small region inside a large quantum many body system,  among all conserved quantities,  only few seem to be relevant. 
For example when a generic  quantum many body system locally relaxes, it does it to a thermal state and thus the only \emph{relevant} conserved quantity is the expectation value of the energy.  Indeed the  Gibbs ensemble (or thermal state) is formally obtained by maximizing the entropy at  fixed  value of the energy \cite{jaynes.57,landau_statistical_1980,pathria_statistical_1996}. Exactly solvable systems, can still locally equilibrate, but to more complex ensembles obtained  by maximizing the entropy subject to the constraints arising from the conservation of all relevant  quantities. It is still unclear in general how to identify the relevant conserved quantities, but in the cases where they are known,  the ensembles that describe the equilibrium of small regions are called  generalized Gibbs ensembles (GGE) \cite{balian_microphysics_2007,rigol2007relaxation}. 

Is entanglement one of those relevant conserved quantities? In order to understand this we address the non-equilibrium dynamics arising after 
a {\em quantum quench} \cite{polkovnikov.11}.
The system is originally in the ground state of a certain Hamiltonian $H_0$. One suddenly quenches the   Hamiltonian from $H_0$ to $H$
and observes the subsequent out of equilibrium dynamics. Depending on if $H$ differs from $H_0$ locally (on few sites) or globally (on the whole system) quenches are called \emph{global} or \emph{local}. 
In particular, we characterize  the quench  obtained by {\em splitting}  a critical spin chain into two equal halves. This amounts to turning off at $t=0$ the interaction between the two half chains. This, together with the fact that the evolution inside each of the two halves  is unitary,  implies that the original entanglement between them is conserved during the
evolution. Thus, the initial  correlations between the two halves are expected to
survive along the whole evolution. 

A similar phenomenon  was observed already in the
experiments carried out by Gring and coworkers in Vienna
\cite{gring_relaxation_2012}, where a quasi-1D Bose gas was split into
two halves. The two halves  were subsequently  allowed to evolve independently. After a time
shorter than the expected equilibration time, many of the observables
had relaxed, a phenomenon usually called {\em prethermalization}
\cite{berges2004prethermalization,kollar_generalized_2011}.  After the prethermalization, the evolution was much slower, and compatible with the effects of the heating of the system due to the residual small interactions with the environment. 
Nevertheless, the original almost stationary interference pattern  between the two halves persisted for large times after the prethermalization time. In a truly isolated system this would have been there forever as a consequence of the initial entanglement between the two halves. A truly isolated quantum system,  indeed, conserves the initial entanglement between two systems that are separated and stop interacting. 
While it is  clear, that by splitting the system one initially injects into the system an amount of extra energy that is proportional to the geometry of the splitting (extensive in the Vienna experiment and intensive in the case we consider here) and thus generate the subsequent out-of-equilibrium dynamics, it is not clear what is the role of the conservation of the entanglement in the subsequent equilibration process.

From a quantum information perspective, the key insight
is that not only the entanglement is conserved, but also each of the  individual eigenvalues of the reduced density matrix of any of the two separated regions is conserved. All together they constitute the  entanglement spectrum (ES)
\cite{li_entanglement_2008}. How does this large amount of constraint affect the dynamics? 

In order to address this point, we  compare the non equilibrium dynamics generated by two similar  quenches. We either  split  a critical spin chain into two halves (we will refer to this situation as to the \emph{split quench}), or we  join two critical chains in a larger one (and we will refer to this scenario as to the \emph{join quench}). Both scenarios are local quenches. Initially, in  the bulk, in middle of the two regions that are either split or joined, any correlation function of  local observables (once appropriately rescaled) is the same in the two cases. Also the post-quench Hamiltonian is the same in the bulk for both quenches. We thus say that the two quenches are in the bulk initially ``locally'' indistinguishable. They are clearly distinguishable close to the boundaries of the sub-systems. While the split quench conserves the initial correlations between the two halves, the join quench does not since the interaction between the two halves allows to distribute correlations among them along the evolution.

The main result that we present here, is that the out-of-equilibrium dynamics and the subsequent relaxation of the bulk of the two systems are distinguishable, and, thus, the presence/absence of conservation of the entanglement spectrum affects the out of equilibrium dynamics of the  system (for related ideas see also Ref. \cite{cazalilla}).

This article is organized as follows. Section \ref{sec:theory}
introduces some general concepts and notation regarding quenches and
thermalization. It also presents our quench protocol and the concrete
model we consider,  the Ising model in a transverse field
(ITF). Section \ref{sec:es} analyses the entanglement structure of the
initial state, and discusses its possible effects on the subsequent
dynamics. In the rest of section \ref{sec:results} we explore the time evolution
of different characteristic magnitudes: the entanglement entropy of
different types of blocks, correlation functions ---both within the
same half and across the split---, and local magnetization. The final
section, \ref{sec:conclusions} is devoted to summarizing the main
conclusions and discussing further work.


\section{The Splitting Quench}
\label{sec:theory}

\subsection{Quenched dynamics and thermalization}

The state $\ket{\phi_0}$ is the ground state (GS) of a certain closed
system described by the Hamiltonian $H_0$. A quench is performed by
changing abruptly the Hamiltonian from $H_0$ to $H$, in such a way
that $\ket{\phi_0}$ ceases to be an eigenstate, and thus undergoes
non-trivial unitary evolution

\begin{equation}
\ket{\phi(t)} = \exp(-iHt) \ket{\phi_0}.
\end{equation}
 $A$ is a (small) region of the system, containing $r\ll N$ spins,
described by the reduced density matrix $\rho_A(t) =\Tr_B
\ket{\phi(t)}\bra{\phi(t)}$, where $B$ is the complement of $A$. Under
certain conditions \cite{popescu_entanglement_2006,gogolin_11,
  riera_thermalization_2012,masanes_complexity_2013}, the limit
$\bar\rho_A \equiv \lim_{t\to\infty} \rho_A(t)$ exists, i.e., for large
enough times, the region $A$ equilibrates to a stationary
state. Typically, a certain amount of time-averaging is necessary in
order to remove small fluctuations.

If, at equilibrium, the state of $A$ is well described by a thermal state, 
it means that, for the equilibration process,  the only relevant conserved quantity is the energy $E$. 
 The  thermal  state is indeed given by $\bar{\rho}_A \simeq \Tr_B\exp
(-\beta H)$, where $\beta$ is  chosen such 
that $\Tr (H \rho_A) /\Tr ( \rho_A)= E$. If the Hamiltonian  is known to commute with  a larger set of relevant local observables, $\set{\braket{H_i}}_{i=1}^K$, the equilibrium
state is a generalization of the thermal state, $\bar{\rho}_A \simeq
\Tr_B \exp (- \sum_{i=1}^K \beta_i H^i)$ and is called a {\em
  Generalized Gibbs ensemble}
\cite{rigol2007relaxation,fagotti_reduced_2013}.

\subsection{The  quench protocol}

\begin{figure}[h!]
\includegraphics[width=\columnwidth]{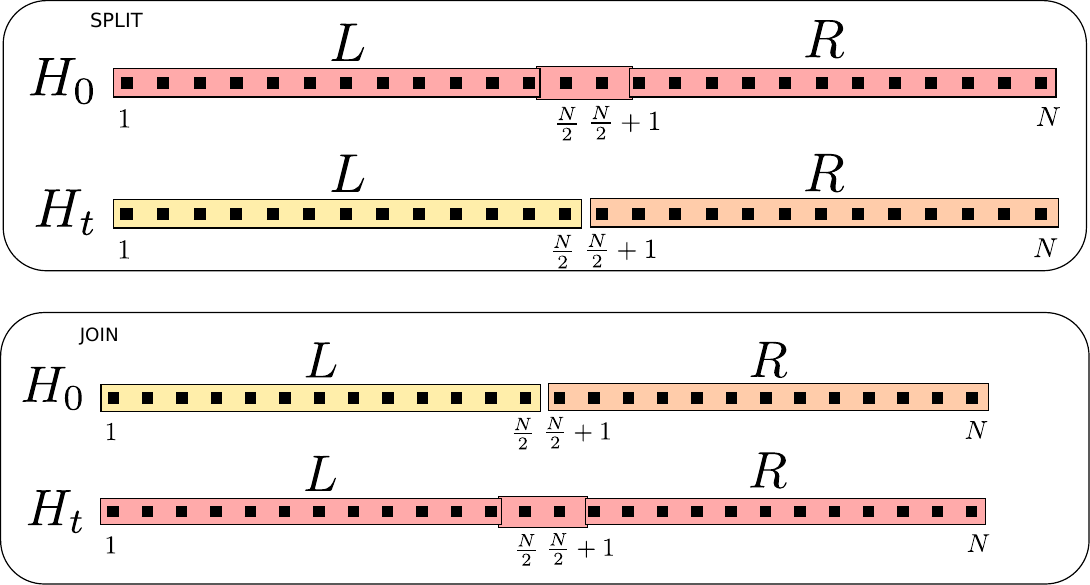}
\caption{Upper panel: \emph{Splitting a spin chain}. After the ground
  state of a spin-chain of length $N$ has been obtained, the
  Hamiltonian is quenched by removing the term which connects both
  halves, effectively splitting them. Lower panel: \emph{Joining two
    spin chains}. The situation is reversed, the ground states of two
  separate chains are quenched by adding the missing term in the
  Hamiltonian which connects them. This provides a reference quench
  for comparison since the two quenches are locally indistinguishable
  and only differ by the presence/absence of the conservation of the
  entanglement spectrum. \label{plot_cadena_def}}
\end{figure}

 Consider a
spin-chain of length $N$, described by a local homogeneous Hamiltonian
with open boundary conditions, $H_0$, which can be formally
decomposed into three terms:

\begin{equation}
H_0 = H_L + H_R + H_{LR}. 
\label{eq:h0} 
\end{equation}
where $H_L$ and $H_R$ act, respectively, on the left and right halves,
and $H_{LR}$ represents the term connecting them. We prepare the
system in its ground state, $\ket{\Omega_0}$, and proceed to
quench the Hamiltonian to

\begin{equation}
H_t \to H = H_L + H_R, 
\label{eq:h1}
\end{equation}
i.e., we remove the connecting term, $H_{LR}$. The upper panel of
Fig. \ref{plot_cadena_def} illustrates the procedure.

The state now evolves as

\begin{equation}
\ket{\phi(t)}= \exp (- i H t) \ket{\Omega_0} 
\label{eq:time_ev}.
\end{equation}
The initial state $\ket{\Omega_0}$ has excess energy with respect
to the ground state of the new Hamiltonian $H$,
$\ket{\Omega_L}\otimes\ket{\Omega_R}$. This excess energy can be
interpreted as the presence at $t=0$ of a finite density of
quasi-particles located at the junction between $L$ and $R$
\cite{calabrese2005evolution}. If  $H$ is a sum of local terms, these
quasi-particles will propagate with a finite speed, giving rise to the
characteristic light-cone effects observed in local quenches
\cite{cheneau_light-cone-like_2012}. Of course, if $H$ contains
long-range interactions, this behavior can differ
\cite{hauke_spread_2013,schachenmayer_entanglement_2013}.

The left  half of the system $L$, at time $t=0$, is described by a mixed state,
obtained by the following reduced density matrix

\begin{equation}
\rho_L(0)=\Tr_R \ket{\Omega_T}\bra{\Omega_T}.
\label{eq:initial_rho}
\end{equation}
that can be diagonalized as,

\begin{equation}
\rho_L(0)=\sum_{\alpha=1}^m \lambda_\alpha \ket{\chi_\alpha}\bra{\chi_\alpha}
\label{eq:rho_diagonalized}
\end{equation}
where $m$ is the Schmidt rank, and the orthonormal $\{\ket{\chi_\alpha}\}$ are
called Schmidt vectors. The subsequent time evolution of $\rho_L(t)$
will be given by

\begin{equation}
\rho_L(t) = U_L(t) \rho_L(0)U_L^\dagger(t)
\label{eq:von_neumann}
\end{equation}
 where $U_L(t)=\exp(-iH_L t)$. An immediate
consequence is that the spectrum of $\rho_L(t)$, the set of
$\{\lambda_k\}_{k=1}^m$, is preserved by the evolution. The Schmidt
vectors, nonetheless, evolve in a non-trivial way, describing a time dependent set of orthogonal vectors. At any later time indeed,

\begin{equation}
\rho_L(t) = \sum_{\alpha=1}^m \lambda_\alpha \ket{\chi_\alpha(t)}\bra{\chi_\alpha(t)}
\label{eq:rho_evolved}
\end{equation}
with the same set of $\{\lambda_k\}_{k=1}^m$ than the one in Eq. \ref{eq:von_neumann}.
It is customary to describe $\rho_L(0)$ in terms of a certain
entanglement Hamiltonian ${\cal H}_L$, such that
$\rho_L(0)=\exp(-{\cal H}_L)$ \cite{li_entanglement_2008}, i.e., as if
it were a thermal state at an effective temperature $\beta=1$. The
entanglement spectrum (ES) is defined to be the set of eigenvalues of
${\cal H}_L$, $\epsilon_\alpha=-\log(\lambda_\alpha)$. Thus, as a consequence of the conservation of the eigenvalues of $\rho_L$, the ES between the left and right parts is also conserved.

How does in general  change the local equilibration after a quench when the ES is conserved? 
In order to address this question we can study the equilibration of a generic mixed state constructed from a set of orthogonal vectors  $\ket{\phi}_{\alpha}$ each appearing with probabilities  $\lambda_\alpha$. We can perform the time evolution for each of the state  individually (it is a linear map) and then reconstruct the 
appropriate mixed state by summing the result with the appropriate probabilities. In the simplest scenario we can assume that
each of the vectors  $\ket{\phi}_{\alpha}$ fulfills the necessary conditions for
thermalization described in
Ref. \cite{riera_thermalization_2012}. Depending on the initial energy of each them $E_{\alpha}$, they  will locally
thermalize to the corresponding temperatures $\beta_{\alpha}$ such that $\Tr (H \exp (\beta_{\alpha} H))/\Tr (\exp (\beta_{\alpha} H)) = E_{\alpha}$. If the $\beta_{\alpha}$ obtained in this way are not sufficiently close,  the final state can not be described by a single
temperature. Similarly, if there are more preserved quantities, the
final state will not be uniquely determined by their initial
expectation values. The final state might, therefore, not be described
by a Gibbs (or generalized Gibbs) ensemble. In other words, the system would
not locally thermalize in the usual sense but it would equilibrate to an exotic ensemble.

Is this non-thermalization likely to occur for the initial mixed state obtained in the split quench? A
generic scaling argument suggests that such temperature mixing is
difficult to achieve. Ground states of gapped 1D Hamiltonians fulfill
the area law of entanglement
\cite{hastings_area_2007,masanes_area_2009}. This implies that the
number of Schmidt vectors saturates with the system size in the
thermodynamic limit. The temperature mixing effect might be more
relevant for a critical initial state, for which the number of Schmidt
vectors scale as a power law of the system size
\cite{vidal_latorre_rico_kitaev}. Still different Schmidt vectors, typically only differ locally so that their initial energies are very similar. We thus do not expect to observe the temperature (or generalized parameter) mixing in our setting. Still, in the results we present, we will  observe some remnants of the fact that the ES is conserved.

In order to clarify the role of the ES-conservation, we will compare
the splitting quench with the joining quench of the same spin-chain,
as illustrated in the lower panel of Fig.~\ref{plot_cadena_def}. In
this last case, one first obtains the ground state of $H_t$
in Eq. (\ref{eq:h1}) and then quenches the Hamiltonian by adding the connecting term
$H_{LR}$, i.e.: applying $H_0$. This  effectively joins the two independent
chains. This case has been addressed both at criticality and away from
it using several techniques, which range from conformal field theories
(CFT) to free fermions \cite{c&c2,dubail,peschel07}.

Since both quenches are locally described by the same Hamiltonian, and the correlation functions of any local operator, in the bulk of the initial states, are  indistinguishable,  we
might expect that the difference between the corresponding out-of-equilibrium evolutions  should be negligible far
away from the division between $L$ and $R$. We will show that this is
not the case, and the two quenches produce substantially different
states, both globally and ---more interestingly--- also locally.

\subsection{The critical Ising chain}

As a prototypical example, let us consider the Ising model in
transverse field (ITF), a simple integrable one-dimensional
spin-chain,

\begin{equation}
H_0 = - \sum_{i=1}^N \left[ \sigma^x_i \sigma^x_{i+1} + \Gamma \sigma^z_i\right]. 
\label{eq:ising}
\end{equation}
where $i$ ranges over the $N$ sites of a 1D lattice and $\sigma^x$ and
$\sigma^z$ stand for the Pauli matrices. The model presents two
phases: a $X$-polarized phase for $\Gamma<1$ and a $Z$-polarized phase
for $\Gamma>1$. They are separated by a second-order phase transition
at $\Gamma_c=1$, where we will perform all our calculations. The ITF
can be rewritten as a free-fermion model via a Jordan-Wigner and a
Bogoliubov transformation \cite{lieb1964two}

\begin{equation}
H_0 = \sum_k \epsilon_k \left(  \eta^ {\dagger}_k \eta_k -\frac{1}{2} \right) \label{eq_ising_free_fermions}
\end{equation}
with $\eta^\dagger_k$ and $\eta_k$ following the usual anticommutation
relations. The model is, therefore, integrable, and all its conserved
quantitites can be expressed as a function of the mode occupations
$n_k$ \cite{fagotti_reduced_2013}:

\begin{equation}
n_k =\bra{\Omega_T} \eta^ {\dagger}_k \eta_k \ket{\Omega_T}. 
\label{eq:mode}
\end{equation} 

The low-energy physics of the ITF model close to the phase transition and its out-of-equilibrium dynamics can also be described using CFT \cite{c&c2,dubail,fagotti2008evolution}. 

In this work we have studied the two quenches via both
free-fermion techniques \cite{torlai_dynamics_2013} and the \emph{time
  evolving block decimation} (TEBD) method, based on matrix product
states (MPS) \cite{vidal_efficient_2003,vidal_efficient_2004}. MPS
techniques have the advantage that they can be extended to both
interacting models and non-integrable models.



\section{Numerical results}
\label{sec:results}

We first analyse the relation between the Schmidt vectors of the initial state  and the eigenvectors of the post quench Hamiltonian $H$. 
This  gives us the opportunity to understand better the possible connections between the conservation of the entanglement spectrum and the long-time equilibrium regime. As we have discussed,  the distribution of the expectation value of the energy and all relevant conserved quantities  taken on the set of the Schmidt vectors are the ultimate quantities that determine if the system equilibrates to a well defined GGE ensemble or not.

We then will proceed to a more traditional  characterization of the states resulting
from the split quench focusing on two types of properties, global and
local. Among the global properties, we will consider the entanglement 
entropy of different types of blocks and large-distance
correlation functions. The local properties are characterized by studying   the expectation
values of local operators.

The entanglement  entropy of a block $A$, with reduced density matrix
$\rho_A$, is defined as

\begin{equation}
 S_A = - \textrm{tr} \rho_A \log \rho_A. \label{eq:ent_entropy}
\end{equation}
We consider both the case in which $A$ is completely contained in one of the two blocks (say $L$) and when it is shared in between $L$ and $R$, see Fig. \ref{plot_cadena_blocks}. 

With respect to correlation functions, we will evaluate the two-point correlation function of the order parameter, defined as

\begin{eqnarray}
C(r_1,r_2,t)&=&\langle\varphi(t)|\sigma_x(r_1)\sigma_x(r_2)|\varphi(t)\rangle \label{eq:formu_correlator}\\
& -&\langle\varphi(t)|\sigma_x(r_1)|\varphi(t)\rangle\langle\varphi(t)|\sigma_x(r_2)|\varphi(t)\rangle. \nonumber
\end{eqnarray}
In particular, we will consider distances $|r_1-r_2|$ scaling with the size of the system, $|r_2-r_1|\propto N$, in order to study the thermodynamic limit.
Since both the splitting and the joining quenches break the translation invariance explicitly, we will consider separately the cases in which both $r_1$ and $r_2$ are on the same side  with respect to the splitting point, or when they lie in different sides.

\begin{figure}[h!]
\includegraphics[width=\columnwidth]{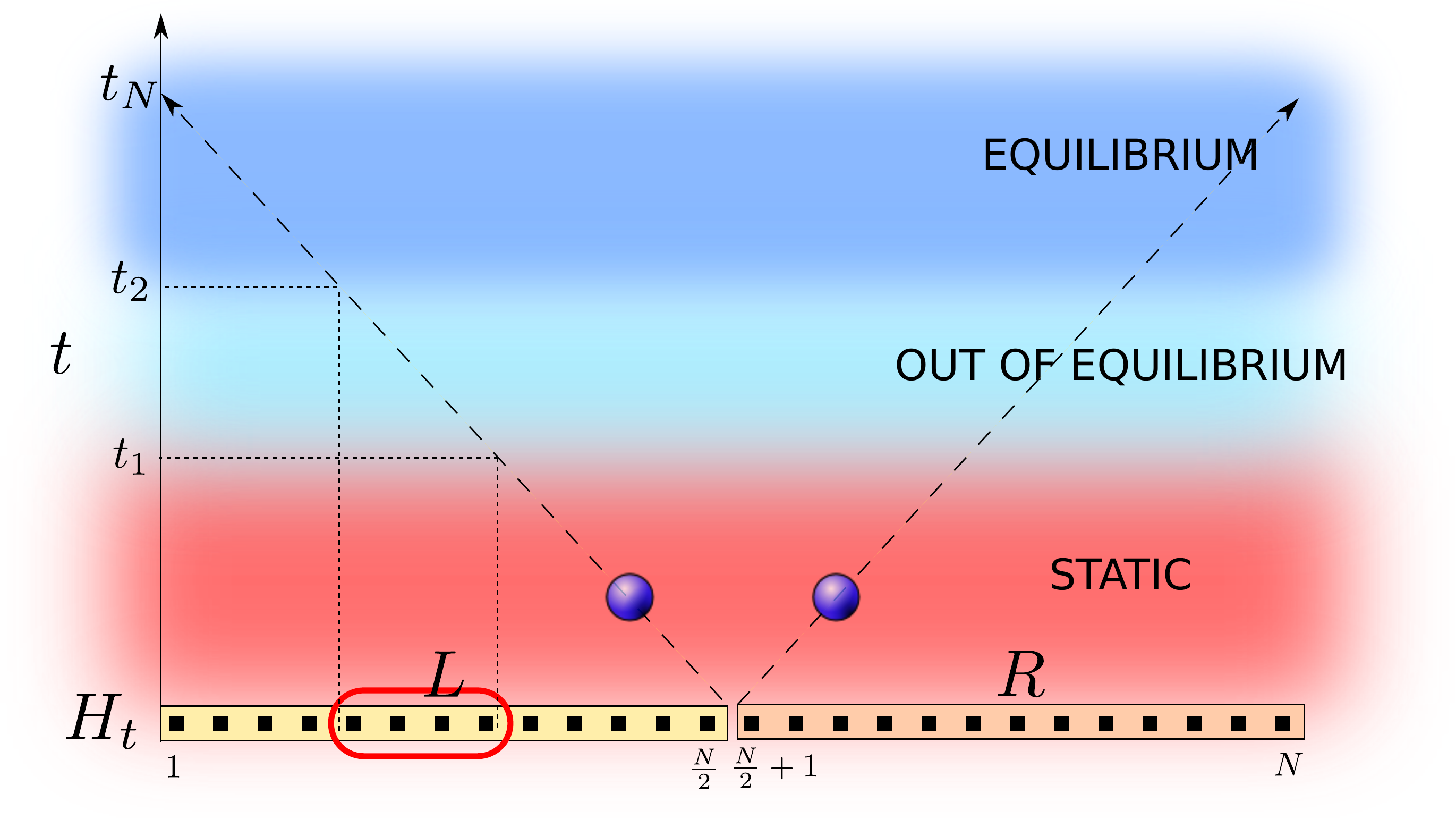}
\caption{\emph{Relevant time scales:} During the out-of-equilibrium evolution which follows the split quench, the region $A$ is characterized by three different regimes as a consequence of the finite speed at which the quasi-particles created at the junction between $L$ and $R$ radiate. A \emph{static} regime, lasting up to time $t_1$, in which the behavior is almost unchanged. This time $t_1$ is indeed the time necessary for the fastest quasi-particle to reach the region. After $t_1$, $A$ experiences an \emph{out-of-equilibrium} regime up to a certain $t_2$, the time needed by the slowest quasi-particles to travel through $A$ and abandon it. From $t_2$ up to $t_N$, we observe \emph{the equilibration} of the region. At $t_N$, the fastest quasi-particles bounce back from the boundaries so that finite size effects start to play a dominant role. \label{fig:time_scales}}
\end{figure}

All those properties will be studied as a function of time. We will also focus on the equilibrium regime which emerges after the transient out-of-equilibrium dynamics. Although global properties, {as discussed in the introduction,} do not equilibrate, in a local quench we can still observe  equilibration of an extensive region $A$ that nevertheless should be separated from the boundaries. Indeed, in our simulations, we always observe three different regimes: (i) the static, (ii) the out-of-equilibrium regime and (iii) the equilibration regime, see Fig. \ref{fig:time_scales}. This fact is well understood by the approximate picture  of the radiation of quasi-particles introduced by Cardy and Calabrese \cite{calabrese2005evolution}. Indeed, through a local quench, one typically populates all single quasi-particle momentum states with equal probability, which then propagate outwards with finite speed $v$. 

Since our model is described at low energy by a CFT, all pseudo-particles propagate with the same speed at first order in $1/N$. In this paper we will not address the  corrections to this picture, that i.e. for long times, are responsible for the spread of the pesudo-particles and thus spoil the periodicity of the dynamics.  

The out-of-equilibrium evolution of a region $A$ which lies at a distance $d$ from the interface between $L$ and $R$ will start after a time $t_1 \approx d/v$. For earlier times, a \emph{static} regime is observed. Eventually, at a time $t_2>t_1$ the slowest particles leave the region and the \emph{equilibration} regime begins. Due to the finite size of the chains, if we wait for a large enough time $t_N \gg t_2$, the quasi-particles will bounce back at the boundary and return to the region $A$, thus making the system depart from equilibration. Thus, we will search for the equilibration regime in the time window $t_2 \gg t \gg t_N$, which depends on the distance from $A$ to the boundaries and the velocities of the different types of quasi-particles. For times $t\gg t_N$ the behavior of the system is plagued with finite-size effects, which we want to avoid since we are interested in the thermodynamic limit. This implies that we can consider at most times of the order of $t_N$.

When addressing local properties, the static regime is followed by a very fast relaxation which leads to the equilibration regime. This can be readily explained assuming that times $t_1$ and $t_2$ coincide.

\subsection{Schmidt vectors of the initial state}
\label{sec:es}
\begin{figure}[h!]
\includegraphics[width=\columnwidth]{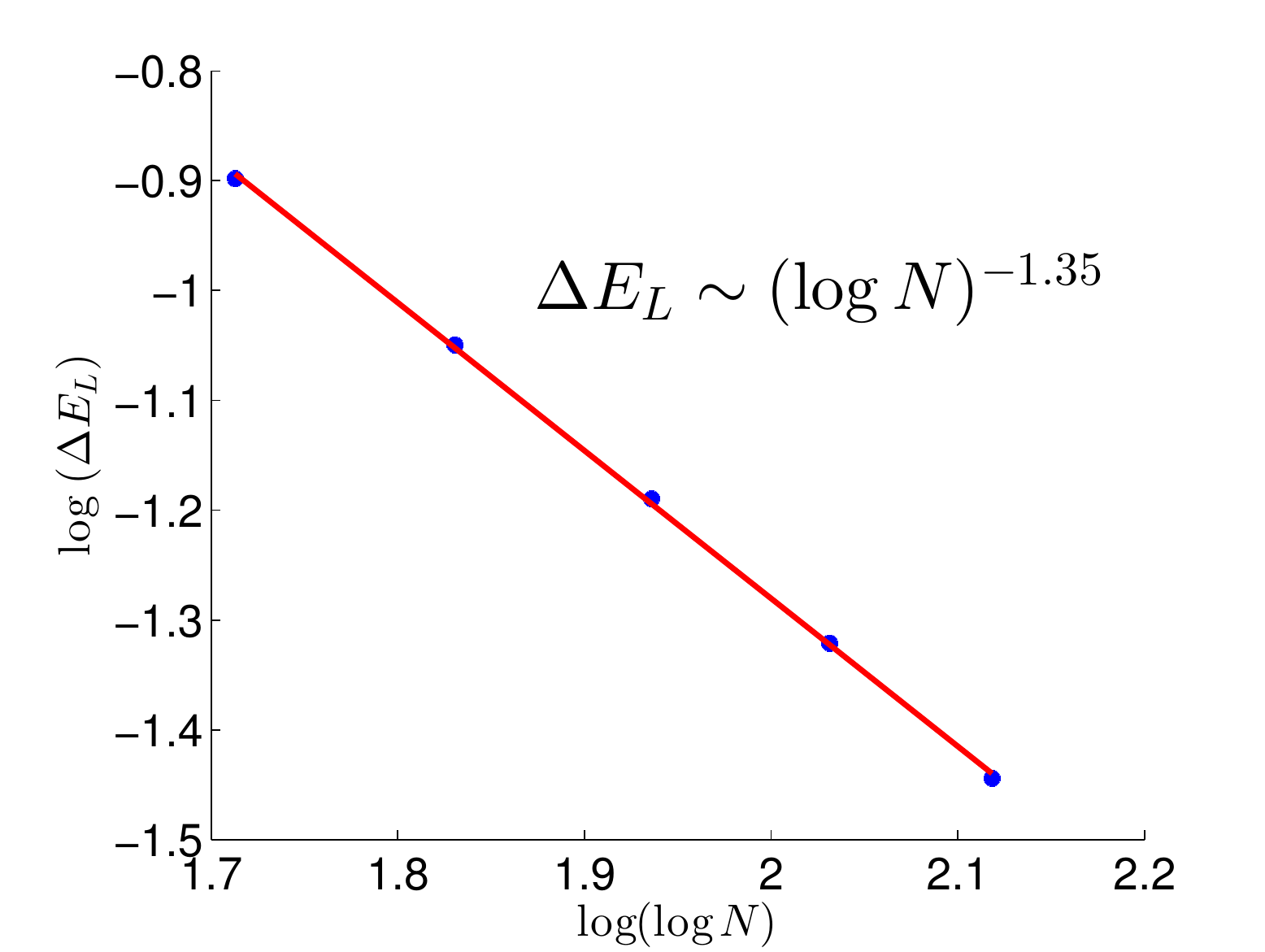}
\caption{Energy gap $\Delta E_L$ between the first two Schmidt vectors
  of the left half of a critical Ising chain as a function of the
  chain length. The data suggest that the gap closes as a power of the
  logarithm of the system size. \label{plot_diff_energy_schmidt}}
\end{figure}

We want to understand the relation between each of the Schmidt vectors and the expectation value of the observables that are conserved during the time evolution.
We start with the energy.  We arrange the eigenvalues of the reduced density matrix
$\rho_L(0)$, $\{\lambda_\alpha\}$, in decreasing order. In this way we can
define an effective {\em energy gap} as

\begin{equation}
\Delta E_L = \bra{\chi_2} H_L \ket{\chi_2}-\bra{\chi_1} H_L \ket{\chi_1} \label{eq:ene_sch_gap}
\end{equation}
We find numerically that $\Delta E_L$ decays as a power of the
logarithm of the system size, as shown in
Fig. \ref{plot_diff_energy_schmidt}. The appearance of a logarithmic
scaling could be related with the results of
\cite{peschel_corner_1987,lauchli_operator_2013,de_chiara_entanglement_2012,alba_2012},
that establish a mapping between the reduced density matrix of the ITF
and the transfer matrix of the corresponding classical model on a
cylinder whose radius grows logarithmically with the size of the
block. 

\begin{figure}[h!]
\includegraphics[width=\columnwidth]{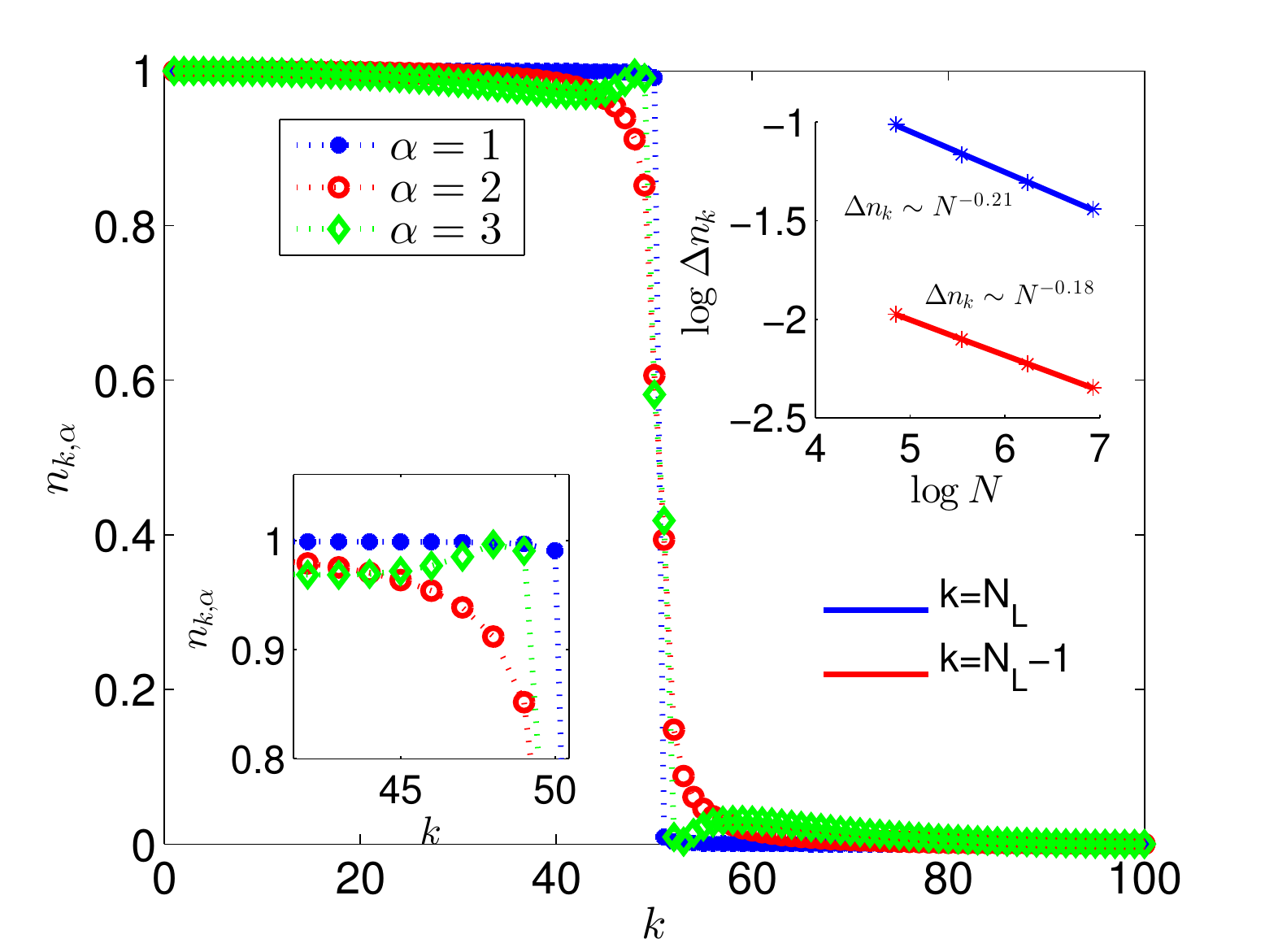}
\caption{Main panel: occupation numbers of the modes of Hamiltonian
  $H_L$ for the first three Schmidt vectors, for chains with $N=100$
  spins. Left inset: the differences in occupation are more pronounced
  around the Fermi energy. Right inset: Those differences scale as a
  power of the system size.}
\label{plot_ocup_schmidt}
\end{figure}

The mode occupations \ref{eq:mode} for the first three Schmidt vectors
are presented in the main panel of Fig. \ref{plot_ocup_schmidt}. They
resemble Fermi-Dirac distribution functions at low temperature, and it
is possible to identify a certain Fermi level that discriminates
between almost fully and almost empty modes. Nonetheless, the
occupations near the Fermi level differ considerably among different
Schmidt vectors, as shown in the left inset of
Fig. \ref{plot_ocup_schmidt}. Those differences decrease slowly as a
power law of the system size (see Fig. \ref{plot_ocup_schmidt}, right
inset).

The Ising model can be mapped to free fermions and thus  all conserved charges are functionally dependent on the mode occupations.
Since for each Schmidt vector the mode occupations are different this 
suggests that this type of quench could provide an example of
equilibration to a strongly correlated state that differs both from the
Gibbs and the Generalized Gibbs Ensembles (GGE), as opposed to what is
expected for standard quenches.

\begin{figure}[h!]
\includegraphics[width=\columnwidth]{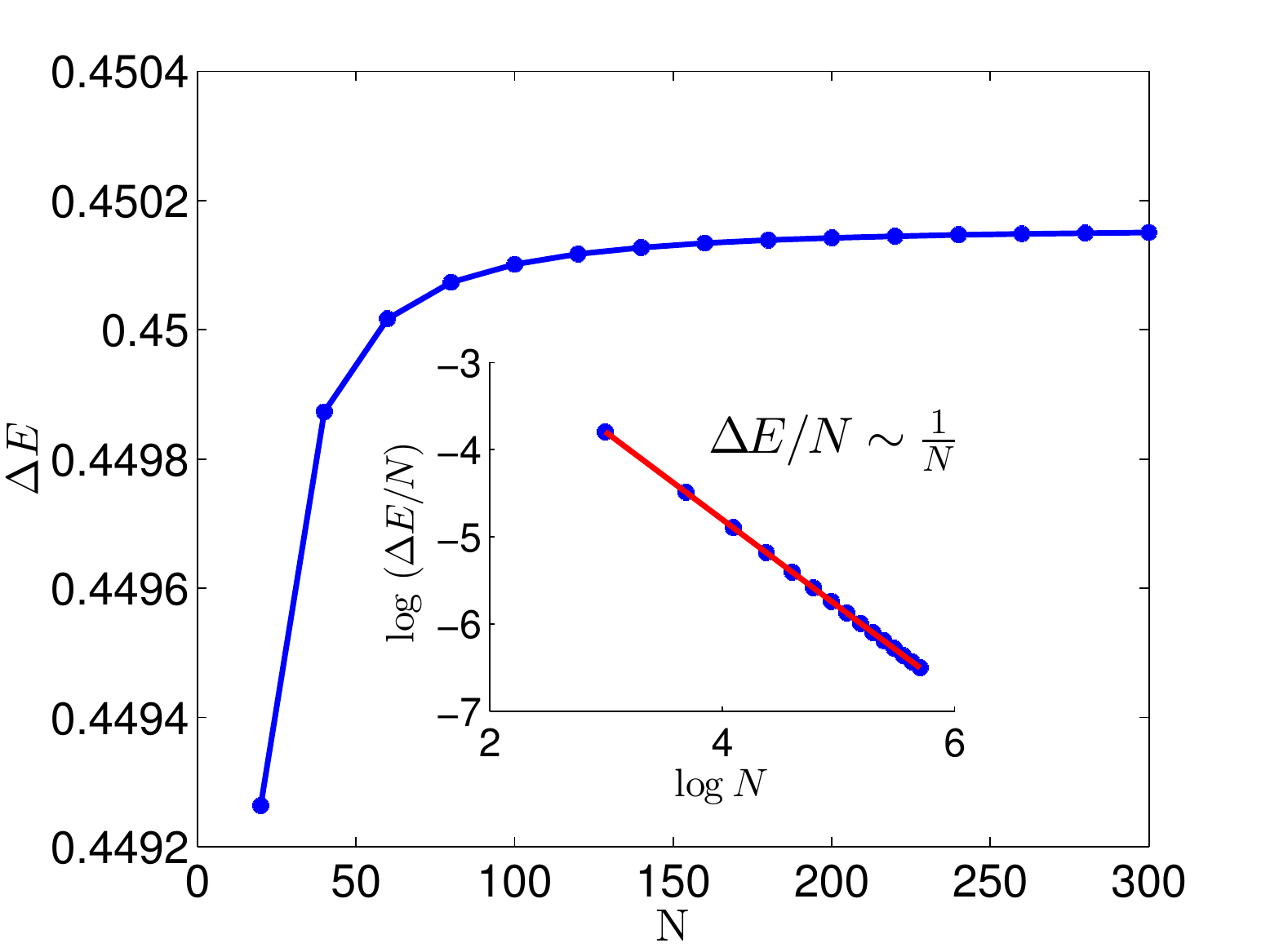}
 \caption{Main panel: Energy fluctuations as a function of the system
   size. We see that local fluctuations (inset) decay to zero as
   $1/N$, as expected in the case of a local quench.}
 \label{fig:variation_ene}
\end{figure}

A more careful quantitative analysis, however, shows that this is not
the case. Indeed, the fluctuations of the energy in the initial state
are not large enough to produce significant effects on the
equilibration state. As shown in Fig. \ref{fig:variation_ene}, those
fluctuations are independent of the system size, so that in the
thermodynamic limit they vanish as $1/N$ (inset).

This is not surprising since, in a local quench, the initial state of
the system does not possess enough energy to equilibrate to a thermal
(or generalized thermal) state characterized by an extensive scaling
of the entanglement entropy of a region. Indeed, in the initial state
of a local quench, the excess energy density with respect to the
ground state scales with $1/N$, and thus it is not surprising that
also its fluctuations scale as $1/N$. This implies that the
equilibrium state of a local quench is very close to a zero
temperature state where, for critical systems, the entanglement
entropy of a region only grows logarithmically with its size
\cite{holzhey1994geometric,callan1994geometric,srednicki1993entropy,
  vidal_latorre_rico_kitaev,c&c}. Still, as we will see in the
following, the conservation of the entanglement spectrum has
non-trivial consequences both on local and global properties of the
system.

\subsection{Entanglement entropy}
\label{sec_entropy}

The time evolution of the entanglement entropy has been computed
analytically in a few selected settings
\cite{calabrese2005evolution,c&c2,dubail,calabrese2007entanglement,fagotti2008evolution},
and numerically in many others, local or global quenches, impurities or
disorder \cite{peschel07,igloi2012entanglement,igloi2012entanglement,
  eisler2008entanglement,lauchli2008spreading,collura2013entanglement,eisler_2014,alba_2014}.

\begin{figure}[h!]
\includegraphics[width=\columnwidth]{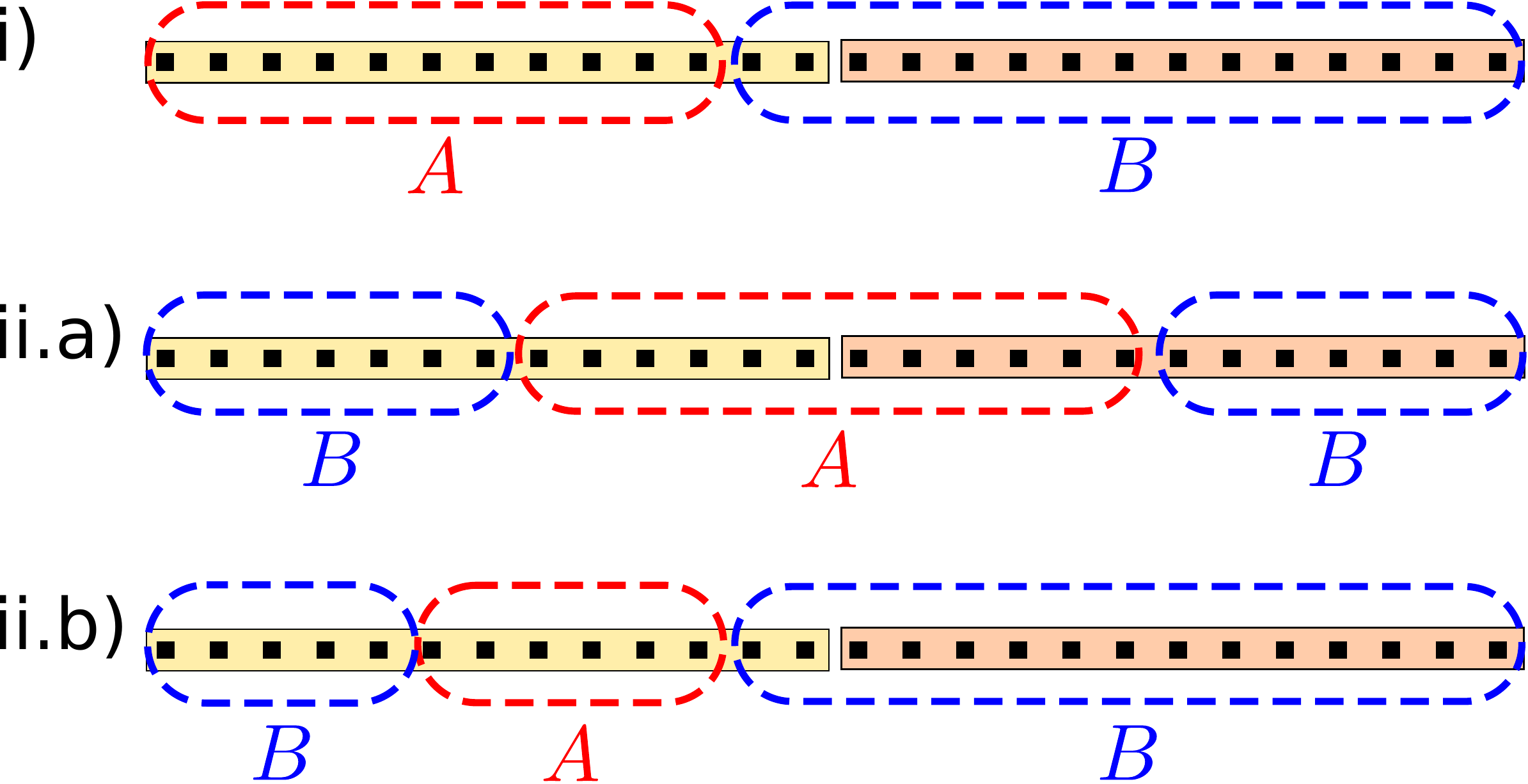}
\caption{The geometrical configuration of block $A$ within the full chain, (i) $A$ has only one active boundary; (ii.a) $A$ has two active boundaries, one in $L$ and the other in $R$; (ii.b) the two active boundaries are both inside $L$. \label{plot_cadena_blocks}}
\end{figure}

In this section we analyze the time evolution of the entanglement entropy of a block $A$ of size $r<N$, as defined in Eq. \ref{eq:ent_entropy}, for two different geometrical configurations (see Fig. \ref{plot_cadena_blocks}), $A$ may have (i) a single active boundary or (ii) two of them. In this second case, the two boundaries may lay on different parts (ii.a) or on the same part (ii.b) of the splitting point.

\begin{figure}[h!]
\includegraphics[width=\columnwidth]{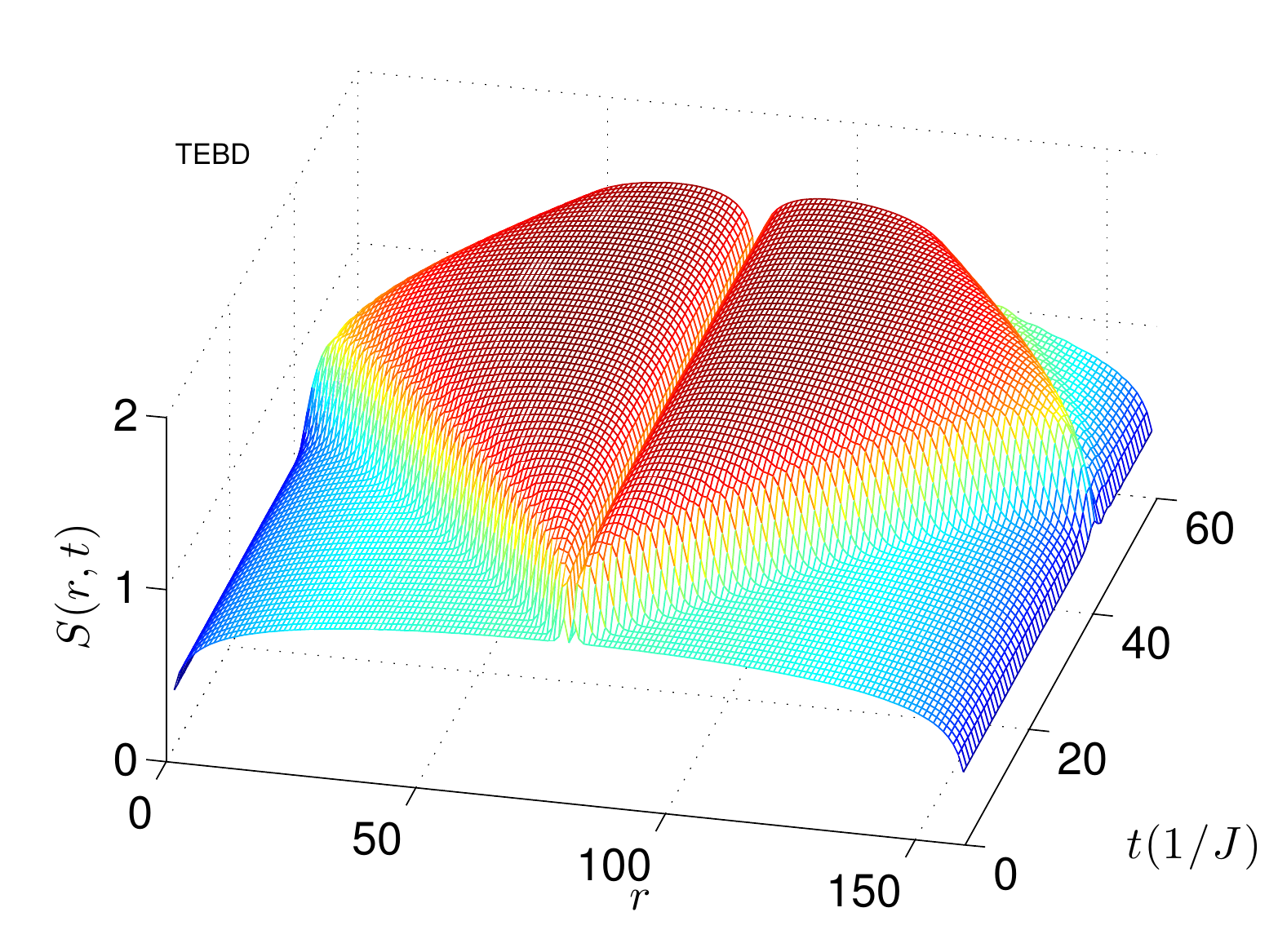}
\includegraphics[width=\columnwidth]{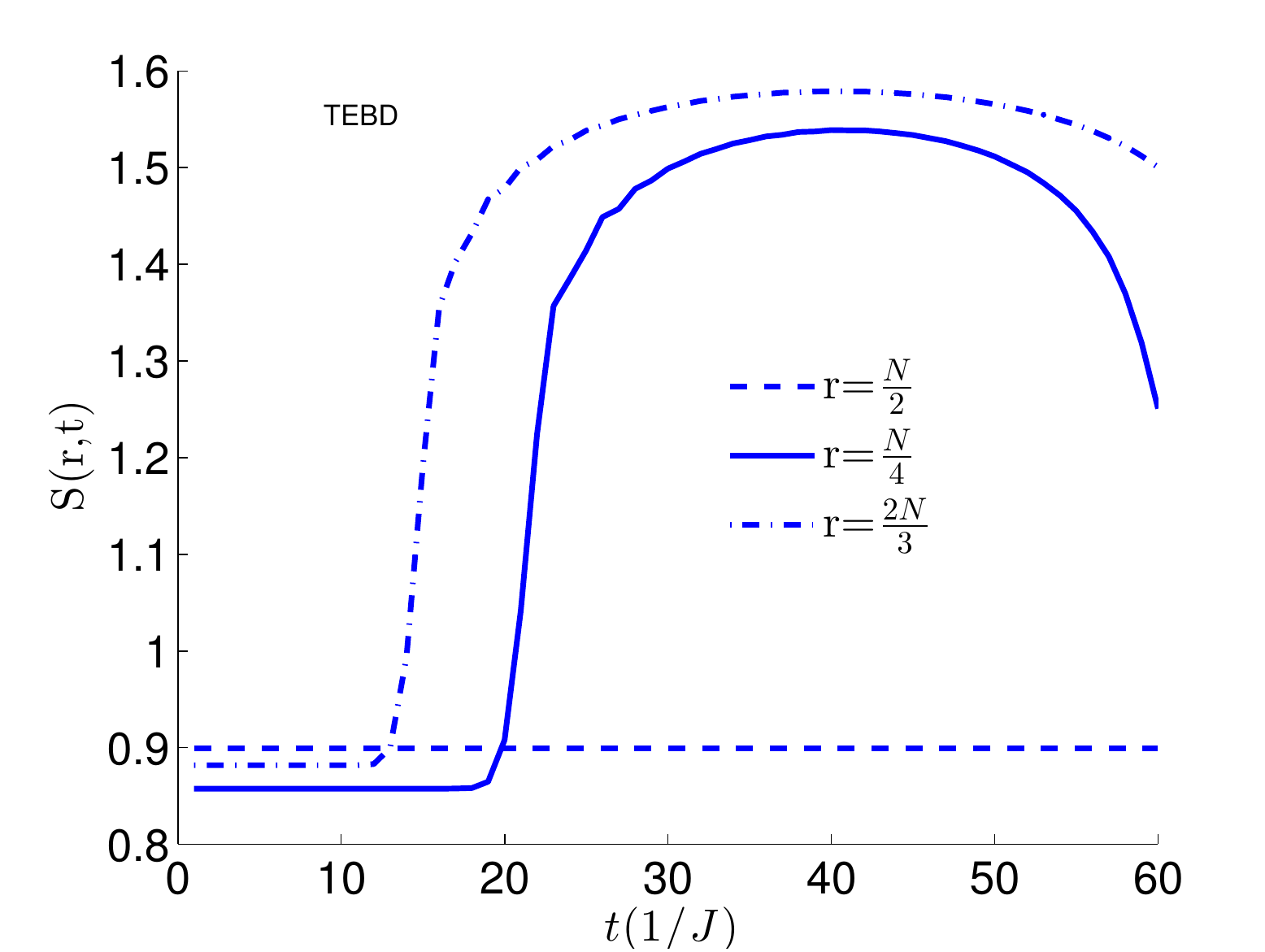}
\caption{\emph{Upper panel:} Time evolution of the entanglement entropy $S(r,t)$ for a chain of $N=160$ after the splitting. \emph{Lower panel:} Time evolution of $S(r)$ for three different values of $r$. When $r=N/2$, entropy remains constant. In the other two cases, entropy starts to grow only after the fastest quasi-particles enter (if $r<N/2$) or leave (if $r>N/2$) the block.}
\label{plot_entr_surface}
\end{figure}

Fig. \ref{plot_entr_surface} is devoted to the analysis of
entanglement in configuration (i). Let $A$ be formed by the leftmost
$r$ sites of a chain with $N=160$ spins, split into two halves. The
upper panel of Fig. \ref{plot_entr_surface} shows the entanglement entropy
$S(r,t)$ as a function of both the size of the block ($X$-axis, marked
$r$), and time in units of $1/J$ ($Y$-axis, marked $t\;(1/J)$). Notice
that, since the entanglement Hamiltonian of the left part, ${\cal
  H}_L$, is a constant of motion, $S(N/2,t)$ is preserved during time
evolution. At $t=0$, $S_r$ presents the characteristic shape of a
critical system: $S(r,0)=\frac{c}{6}\log \left( \frac{L}{\pi }\sin
\left(\frac{\pi l }{L}\right) \right)$
\cite{holzhey1994geometric,callan1994geometric,srednicki1993entropy,vidal_latorre_rico_kitaev,c&c}. But
for further times, a light-cone develops at the $LR$ interface, and
the entanglement entropy only changes when the fastest quasi-particles
generated at the quench cross the active boundary of $A$ as a specific
case of the cartoon sketched qualitatively in
Fig. \ref{fig:time_scales}.

The lower panel of Fig. \ref{plot_entr_surface} shows the time evolution of $S$ for blocks of type (i) and different sizes. $S(N/2,t)$  is constant; $S(N/4,t)$, has a single active boundary that  lies at the left of the splitting point, and thus it presents a stationary behavior for short times followed by  a fast increase when the  quasi-particles reach $N/4$; we also analyze $S(2N/3,t)$, whose active boundary lies to the right of the split point. In this case quasi-particles are radiated from within the region and only contribute to an increasing entropy  when they leave the region. 

\begin{figure}[h!]
\includegraphics[width=\columnwidth]{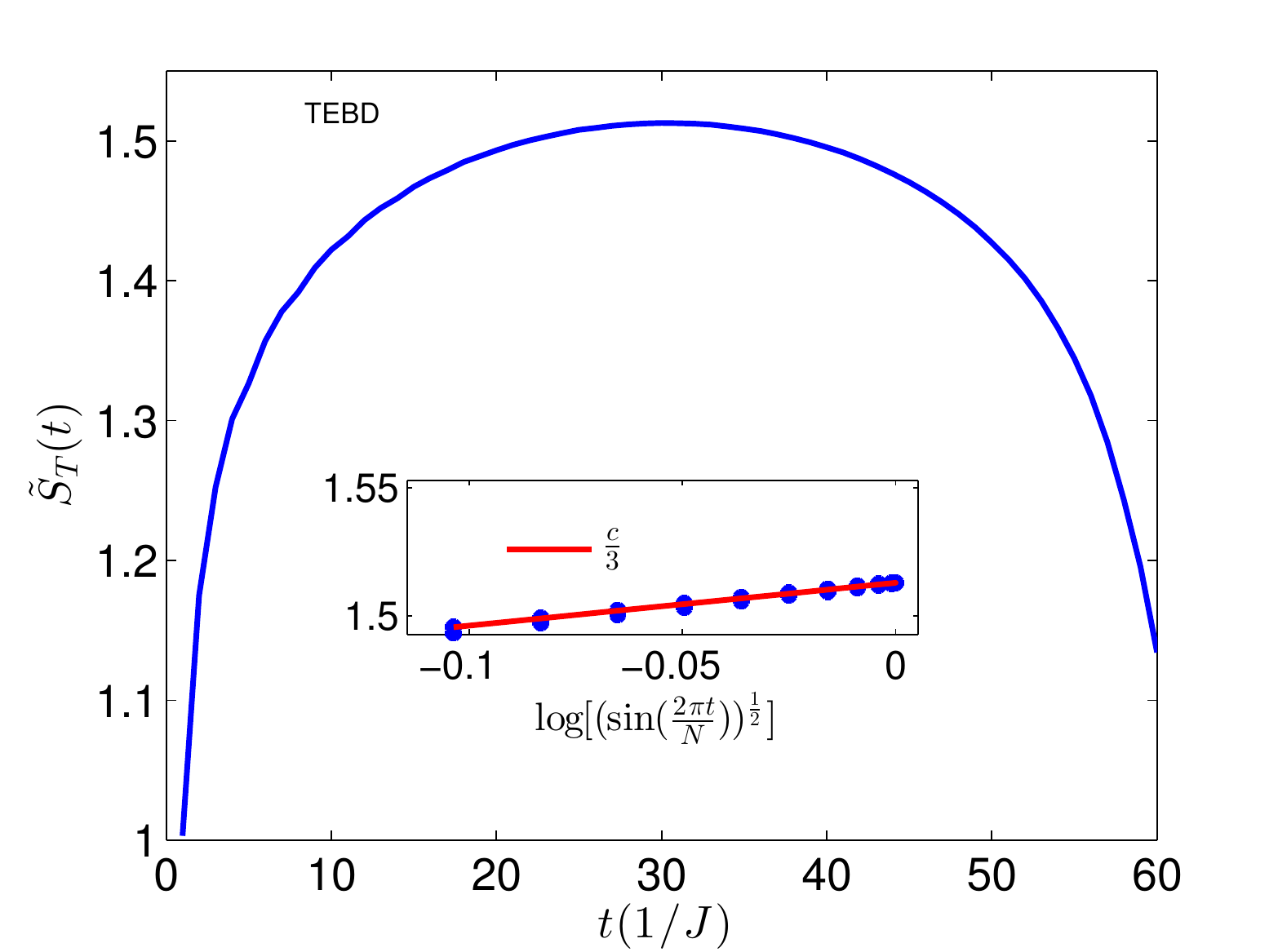}
\includegraphics[width=\columnwidth]{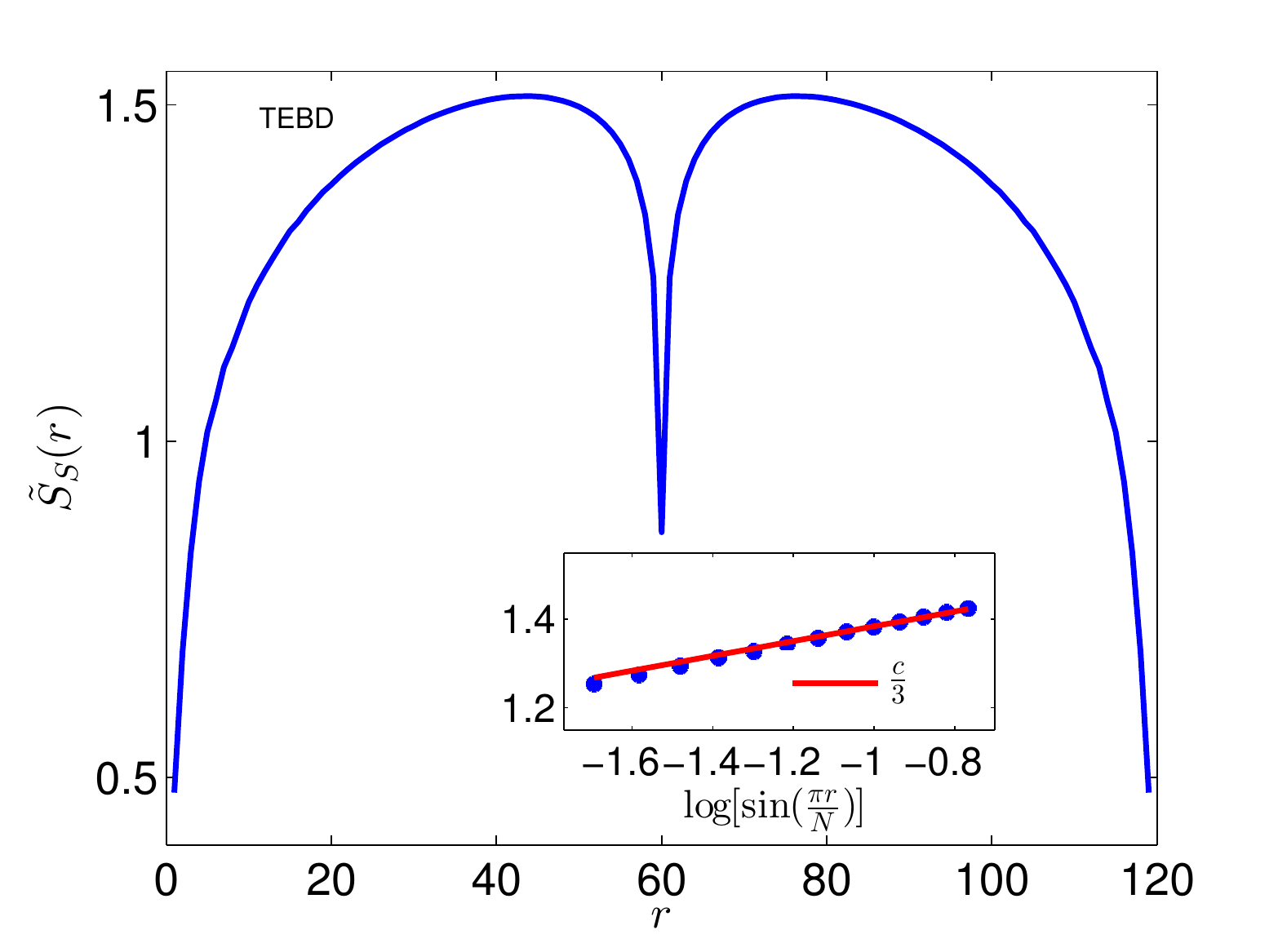}
\caption{\emph{Upper panel:} Time-like projection of the entropy, $\tilde S_T(t)$, after a splitting quench. The inset shows that the growth is compatible with  a logarithmic growth with a pre-factor close to $\frac{c}{3}$, as in the join quench, but with an extra square root inside the logarithm (see Eq. \ref{eq_ent_projection}). \emph{Lower panel:} Space-like projection, $\tilde S_S(r)$. For block sizes very different from $N/2$, it behaves as in the join quench, displaying a logarithmic growth with a pre-factor close to $\frac{c}{3}$. \label{plot_entr_projection}}
\end{figure}

From the space-time diagram of the upper panel of Fig. \ref{plot_entr_surface} we can extract two projections. A time-like projection $\tilde S_T(t)$ is obtained by finding, for each time $t$, the maximal entropy among all block-sizes, and a space-like projection $\tilde S_S(r)$  defined by finding, for each block-size, the maximal entropy achieved along the evolution. After a joining quench, the time-like projection $\tilde S_T$ only grows logarithmically with time \cite{dubail,igloi2012entanglement}

\begin{equation*}
\tilde{S}_T(t)=\frac{c}{3}\log_2\left|\frac{N}{\pi} \left( \sin\frac{\pi v t}{N} \right) \right|+\textrm{cst}
\end{equation*}
where $c=\frac{1}{2}$ is the central charge of the critical Ising chain and $v$ is the quasi-particle velocity. The space-like projection, $\tilde{S}_S(r)$ is described by the same equation, just replacing $t$ with $r$. On the other hand, after a splitting quench, $\tilde S_T(t)$ behaves as

\begin{equation}
\tilde S_T(t)=\frac{c}{3}\log_2\left|\frac{N}{\pi} \left( \sin\frac{\pi v t}{N} \right)^{1/2} \right|+\textrm{cst},
\label{eq_ent_projection}
\end{equation}
where the main difference with the result for the joining quench is the presence of the square root (see upper panel of Fig. \ref{plot_entr_projection}). The space-time projection $\tilde S_S(r)$ presents a cusp at $r=N/2$ which is absent in the joining case. Still, for small $r$, the two cases are difficult to distinguish (see lower panel of Fig. \ref{plot_entr_projection}). The detailed analysis is presented in Appendix \ref{app_entropy_scaling}.

\begin{figure}[h!]
\includegraphics[width=\columnwidth]{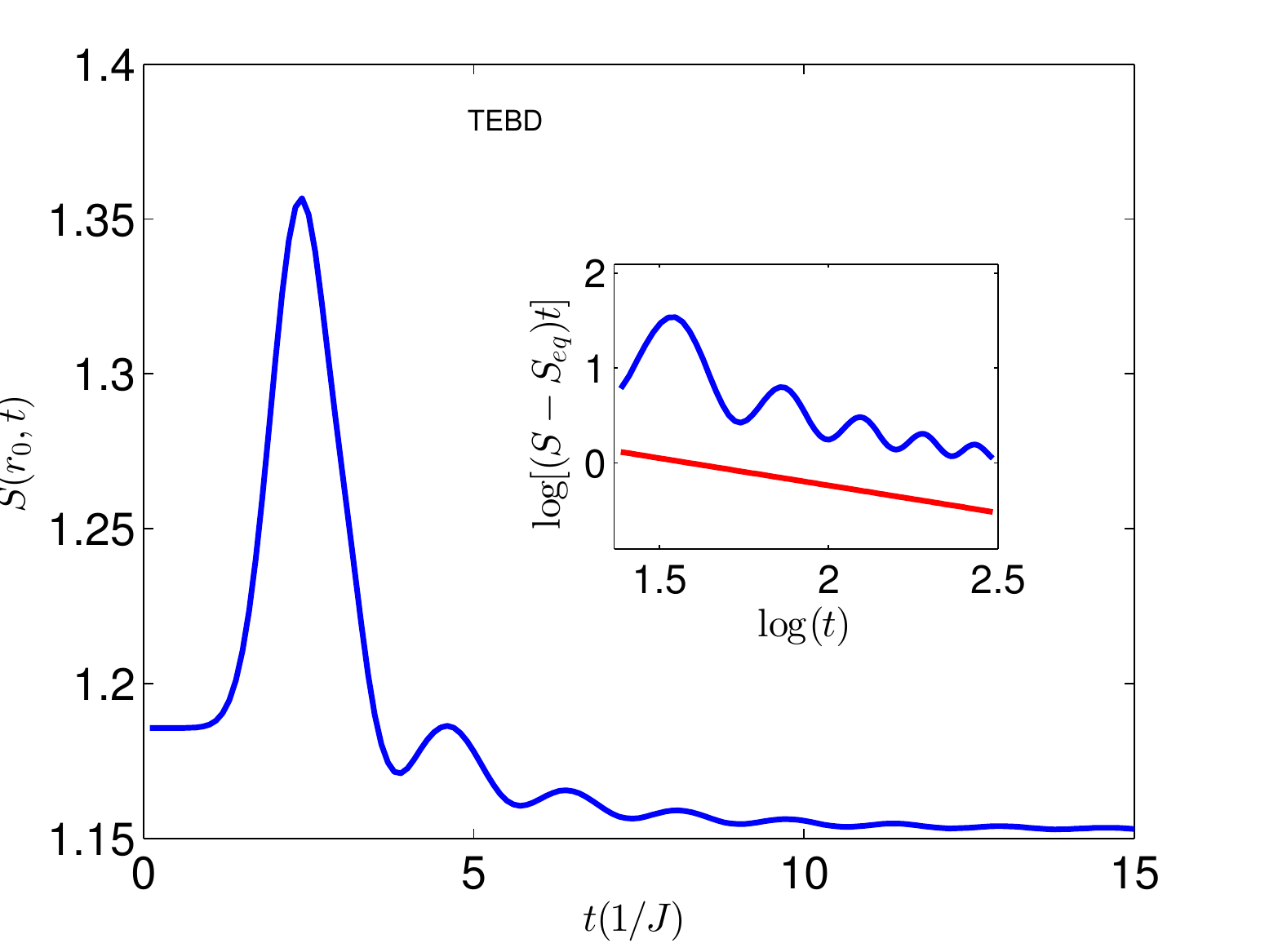}
\includegraphics[width=\columnwidth]{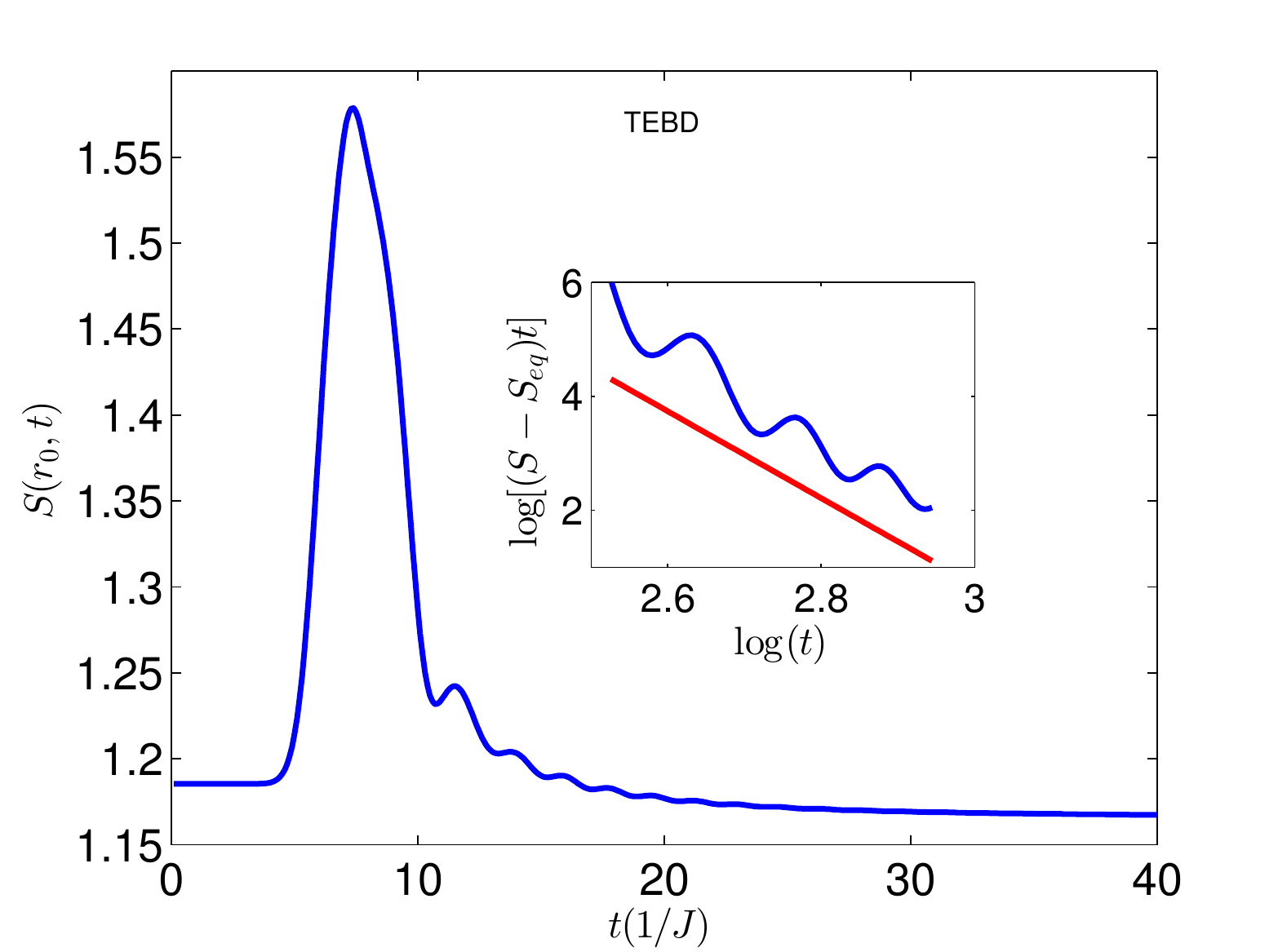}
\caption{\emph{Upper panel:} Time evolution of the entanglement entropy for a block with $r_0=8$ sites centered on the $LR$ interface in a spin-chain of $N=160$ sites. Notice the sudden jump to a maximum value and the slower relaxation towards $S_{eq}$. \emph{Lower panel:} Same evolution, but for a block with $r_0=8$ sites, located at a distance $l=10$ from the interface. In both cases the insets show that the relaxation behavior is compatible with the one of  Eq. \ref{eq_relaxation_entropy}, as in the case of the joining quench, but with faster timescales and super imposed oscillations.}
\label{plot_entr_block}
\end{figure}

The block in configurations (ii) in Fig. \ref{plot_cadena_blocks} have two active boundaries. As discussed previously, we distinguish between blocks (ii.a) which overlap with both parts, which we will place centered on the $LR$ interface, and (ii.b), those which lie totally within one part. In Fig. \ref{plot_entr_block} we consider two blocks of size $r=8$, one of them centered on the $LR$ interface (ii.a) and the other at a distance $l=10$ from it. Notice that the time evolution of the entropy of two blocks presents the three aforementioned stages: static, out-of-equilibrium and equilibrium. In both cases, at large times the entropy converges to a finite value $S_{eq}$.

In the joining quench the relaxation towards $S_{eq}$ is governed by \cite{peschel07}

\begin{equation}
S(r_0,t)=S_{eq}+\frac{\alpha\;\log(t)+\beta}{t},
\label{eq_relaxation_entropy}
\end{equation}
where the parameters $\alpha$ and $\beta$ depen on the distance $l$ between the block and the site of the quench.

 The insets of Fig. \ref{plot_entr_block} unveil a leading behavior  of the same type  as  Eq. \ref{eq_relaxation_entropy} with superimposed  oscillations  and  faster time scales.

\begin{figure}[h!]
\includegraphics[width=\columnwidth]{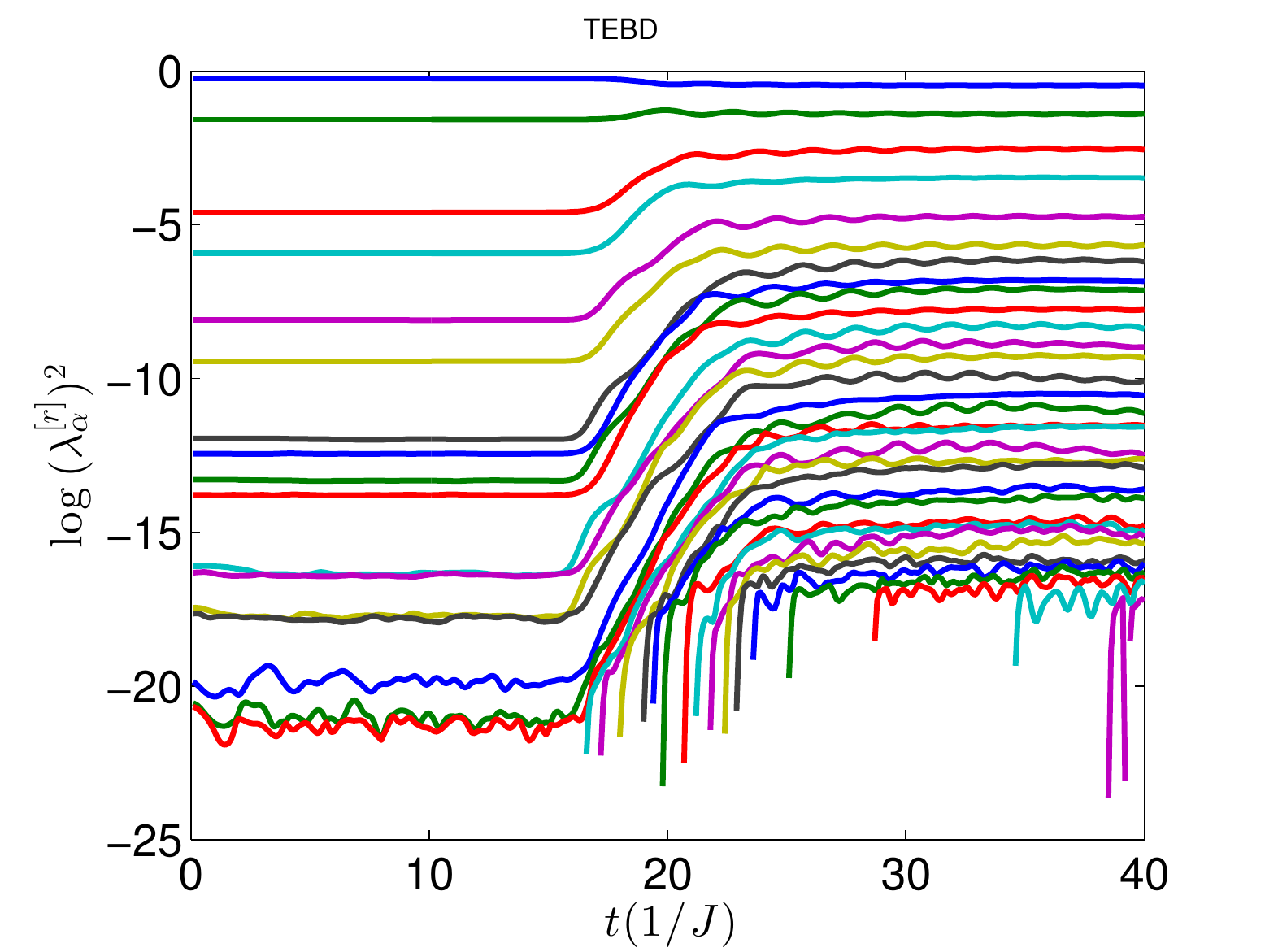}
\includegraphics[width=\columnwidth]{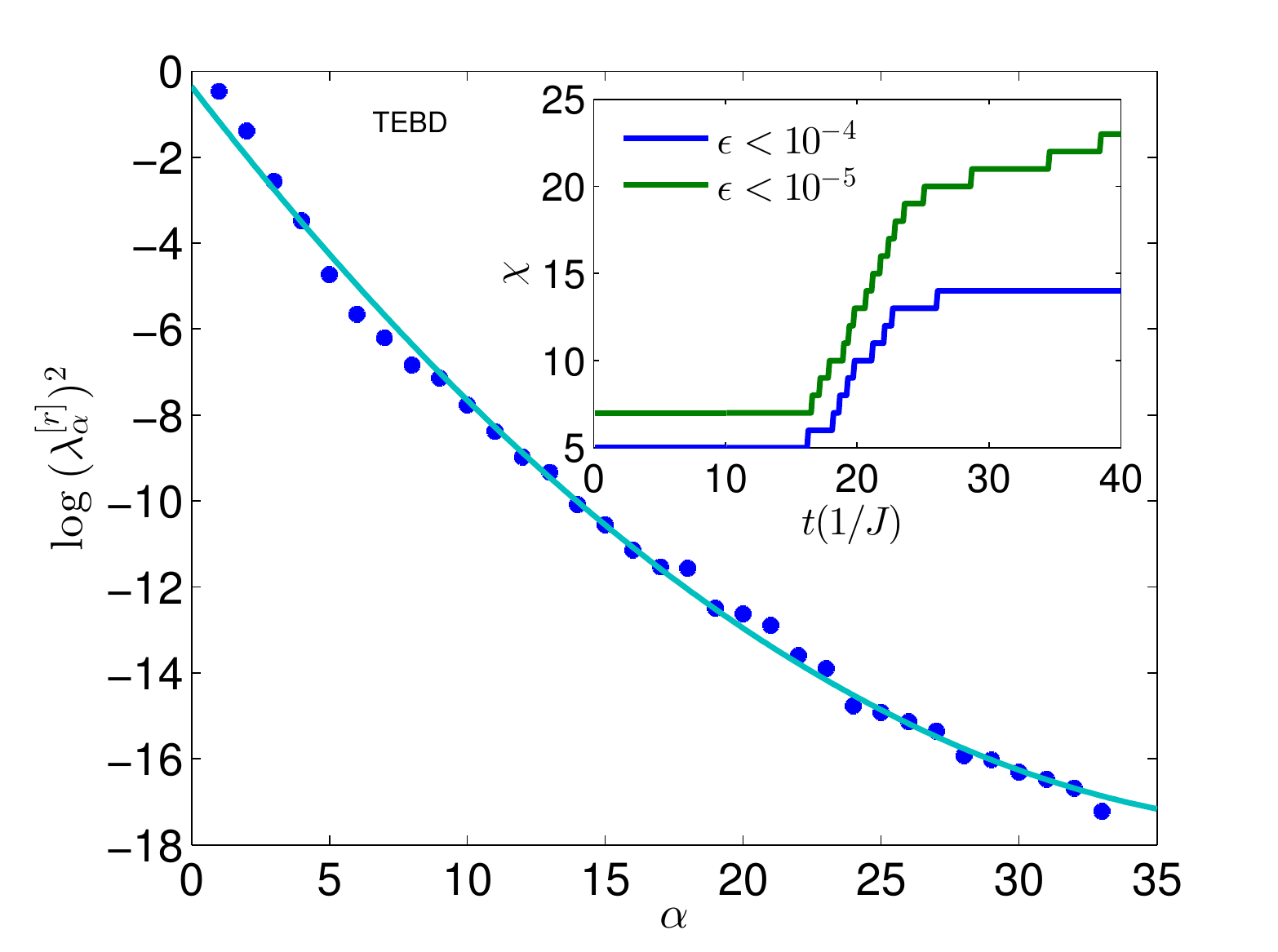}
\caption{\emph{Upper panel:} Time evolution after a split quench of the entanglement spectrum for a block of type (i) with size $N/4$ in a chain of length $N=140$. Notice that only two time regimes are present: static and  out-of equilibrium. \emph{Lower panel:} After the fast increase ($t=40$), the entanglement spectrum decays polynomially, showing that the state is neatly approximable with a few Schmidt vectors, $\chi$. Small values of the representation error are obtained with $\chi\approx 20$, as shown in the inset.}
\label{plot_corre_dif_half}
\end{figure}

The entanglement spectrum of a block in configuration (i) of Fig. \ref{plot_cadena_blocks} presents only  the first two out of the three time regimes, static and out-of-equilibrium. This is a consequence of the fact that  the block extends up to the extreme of $L$ and the quasi-particles never have space to escape. The Schmidt coefficients of the reduced density matrix of a block increase abruptly  when the quasi-particles reach the block, and slower further increase afterwards. During this last regime, the Schmidt coefficients $\lambda^r_\alpha$ decay as a power of $\alpha$ (see lower panel of Fig. \ref{plot_corre_dif_half}), pointing to the possibility of approximating  the state by keeping only a small number of them, $\chi$. The error of this approximation, which is the usual systematic error of MPS-based techniques, is given by

\begin{equation}
\epsilon=1-\sum_{\alpha=1}^{\chi}(\lambda^r_{\alpha})^2.
\end{equation}
The lower panel of Fig. \ref{plot_corre_dif_half} shows also the Schmidt number $\chi$as a function of time required to achieve  two possible desired tolerances $\epsilon$. Notice that the value of $\chi$ increases only moderately during the whole time interval, justifying our choice to use the TEBD algorithm \cite{perales2008entanglement}.

\subsection{Correlation functions}

\begin{figure}[h!]
\includegraphics[width=\columnwidth]{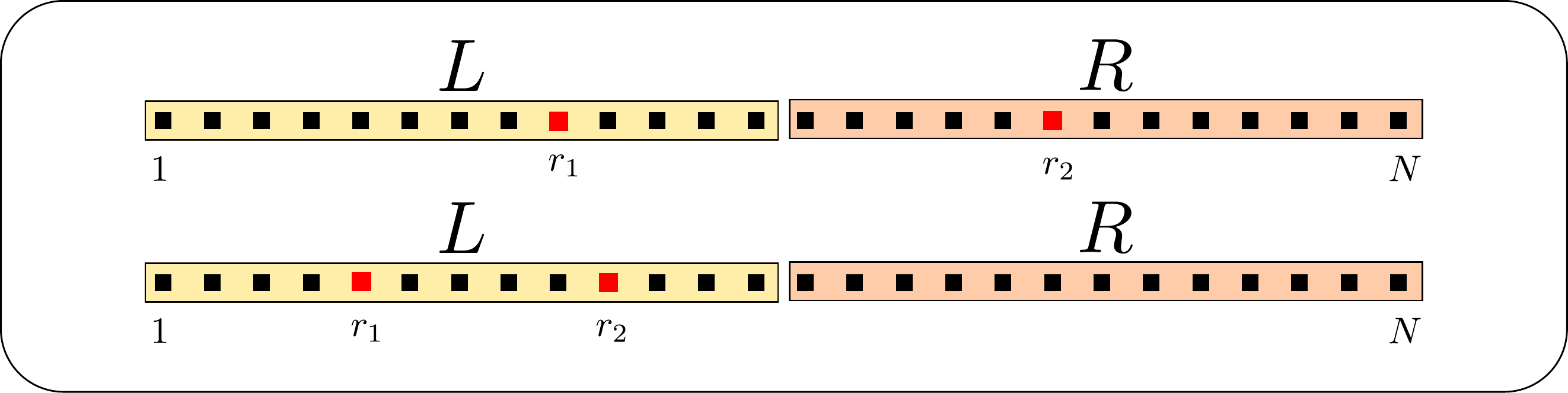}
\caption{\emph{The two-point correlation function of the order parameter,} defined in Eq. \ref{eq:formu_correlator}, is studied in two different configurations,  when the two sites are on different halves (top) and when they lie on the same half (bottom) of the split chain.}
\label{plot_cadena_corr} 
\end{figure}

Let us turn to the time evolution of the two-point correlation functions of the order parameter after splitting the chain, defined in Eq. \ref{eq:formu_correlator}, and compare them with the joining case, which has been studied in detail by several authors \cite{c&c2, divakaran2011non}. As with the entanglement entropy, we will study them in two geometric configurations, shown in Fig. \ref{plot_cadena_corr}, when both sites $r_1$ and $r_2$ are in different halves of the chain (top panel) and when they lie in the same half (bottom).

\begin{figure}[h!]
\includegraphics[width=\columnwidth]{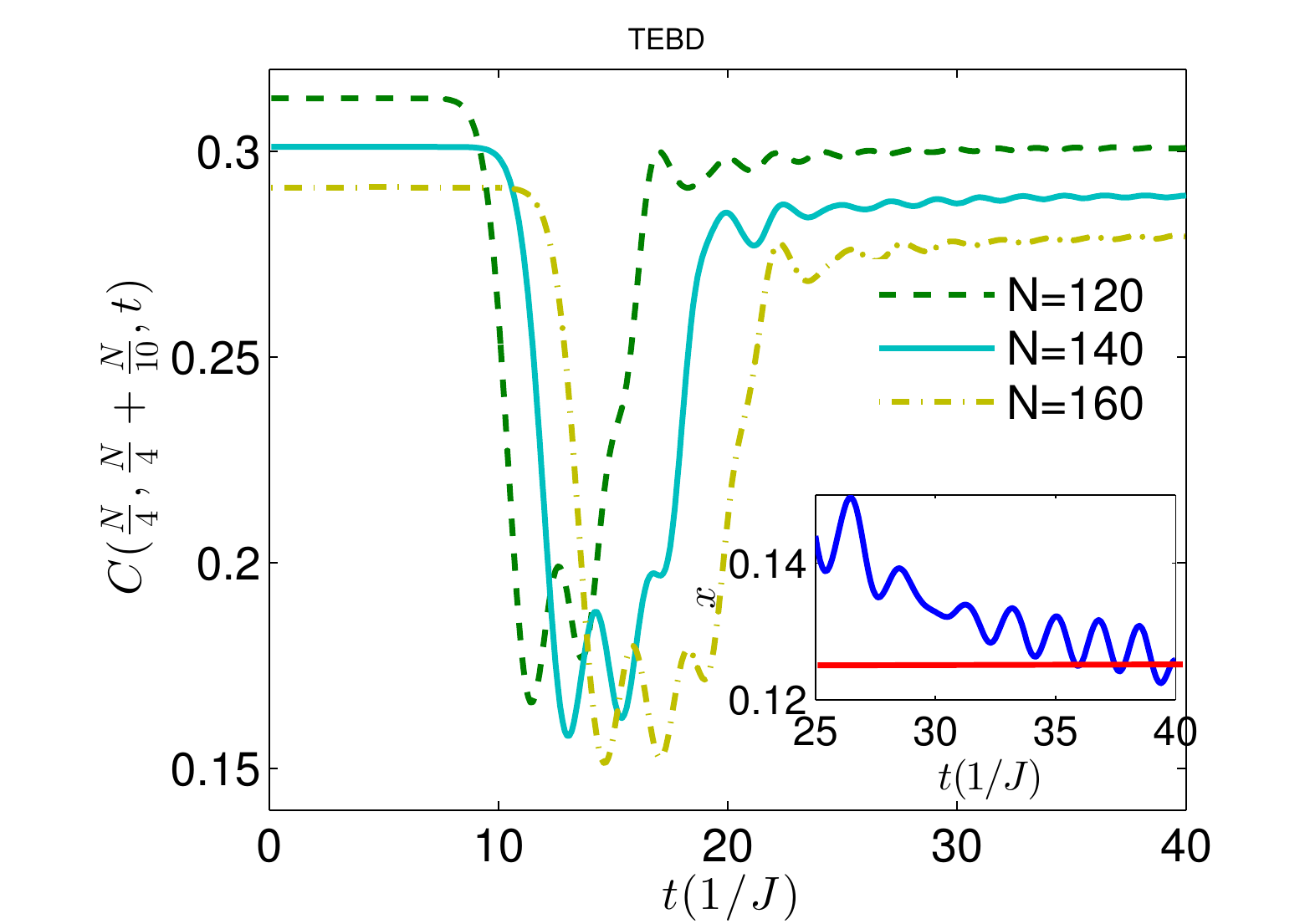}
\includegraphics[width=\columnwidth]{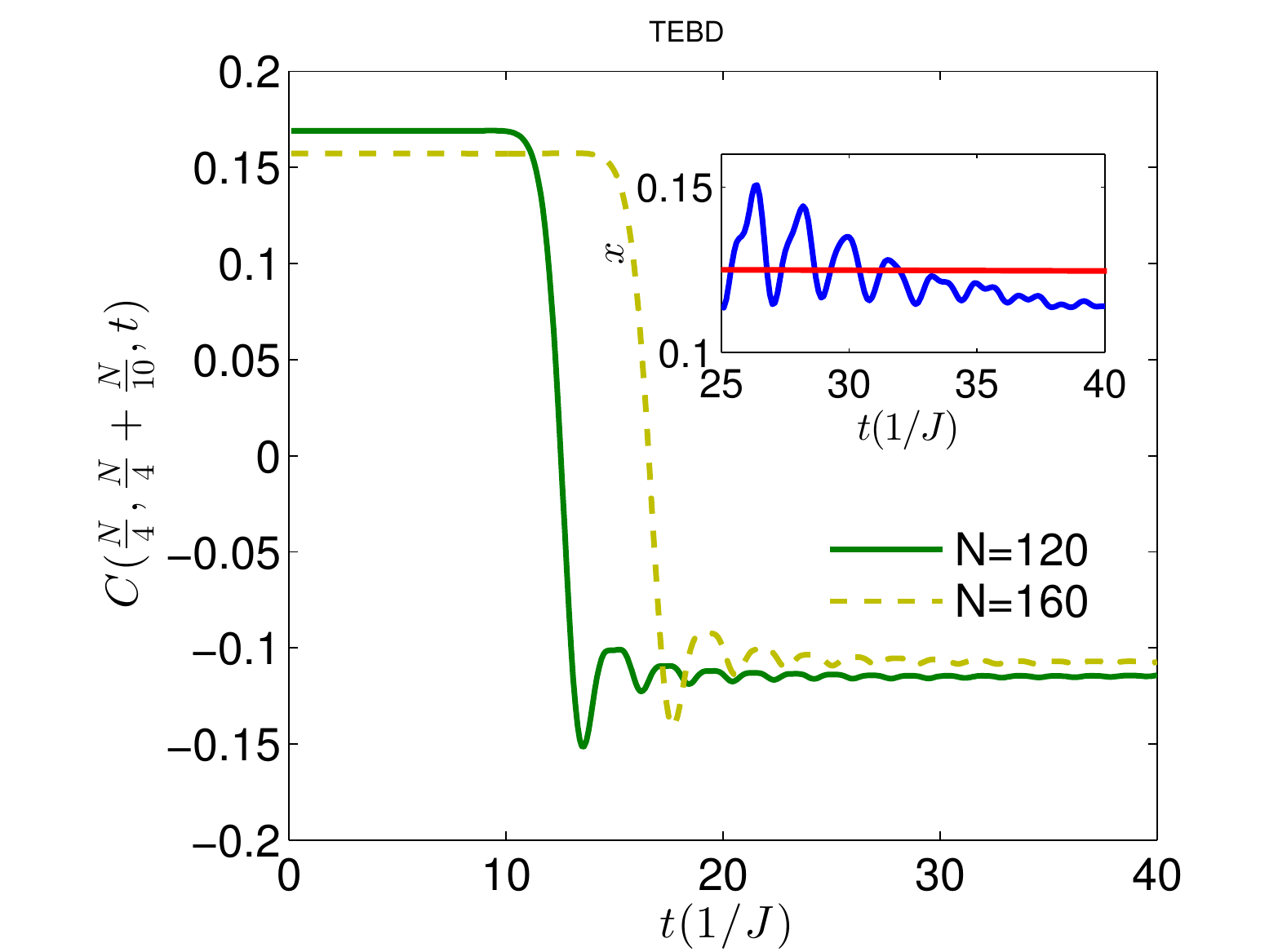}
\caption{\emph{Time-evolution of the correlation functions.} In the upper panel we consider the case when both points are in the same half of the chain. Concretely, we plot $C(N/4,N/4+N/10,t)$ for different chain lengths $N$. In both the static and equilibrium regime we observe a polynomial decay of the correlation as a function of the system size $N$. The critical exponent $x$ governing the decay has been monitored as a function of time during the equilibrium regime, showing oscillatory convergence to the static value $1/8$ (see inset). The lower panel shows the time evolution of the correlator when both points lie in different halves, symmetrically placed with respect to the $LR$ interface. Concretely, we plot $C(N/2-N/10,N/2+N/10,t)$. Interestingly, the equilibrium regime leads to anti-correlation between the two half-chains, whose critical exponent $x$ again converges through some oscillations to $1/8$ (inset).}
\label{plot_corre_same_half}
\end{figure}

In the static regime, since the system is critical, for $|r_2-r_1| \propto N$, $C(r_1,r_2,t)\propto N^{-2x}$ with $x=1/8$ (see Fig. \ref{plot_corre_same_half}, both panels, for short times).

\begin{figure}[h!]
\includegraphics[width=\columnwidth]{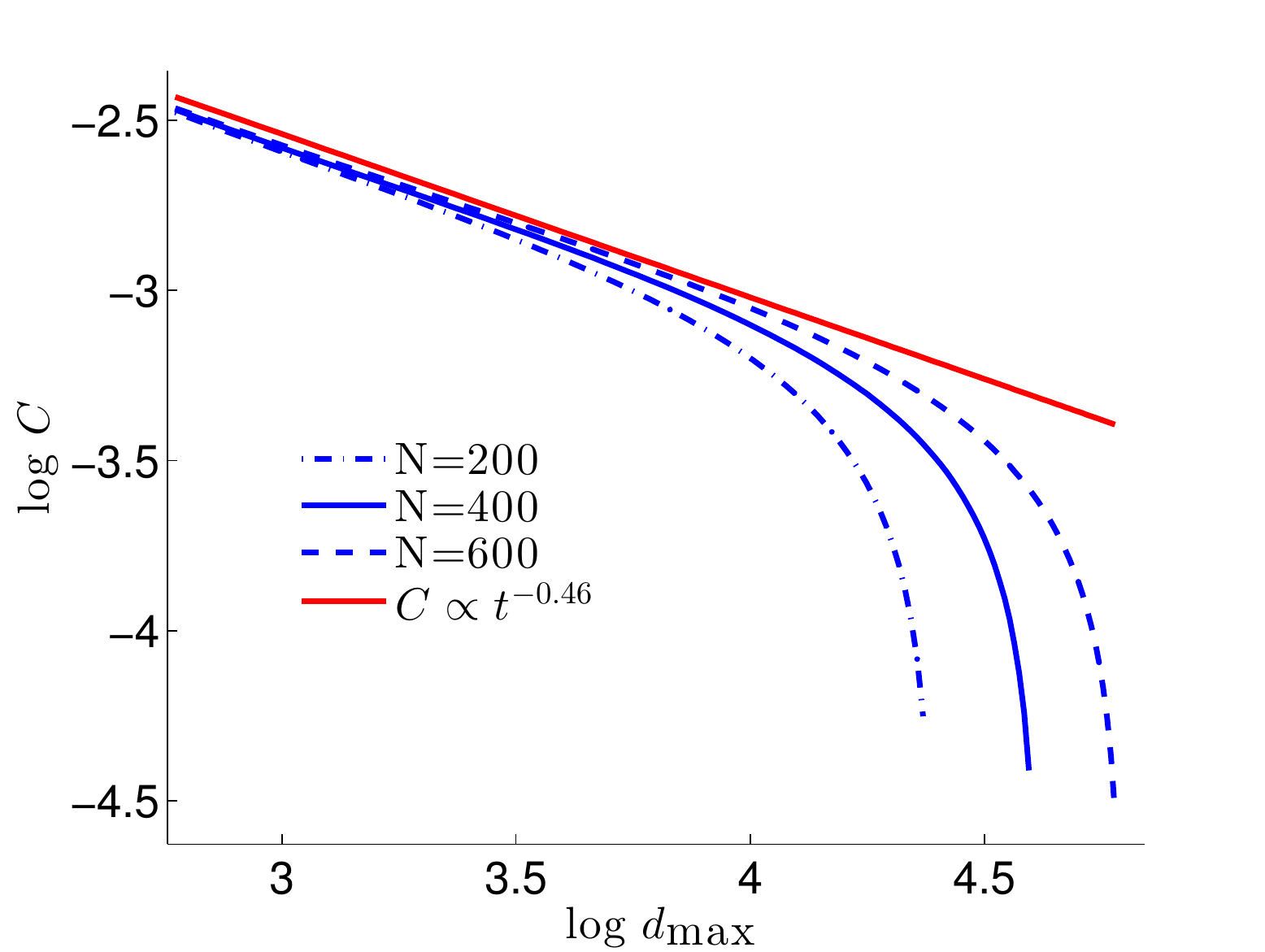}
\includegraphics[width=\columnwidth]{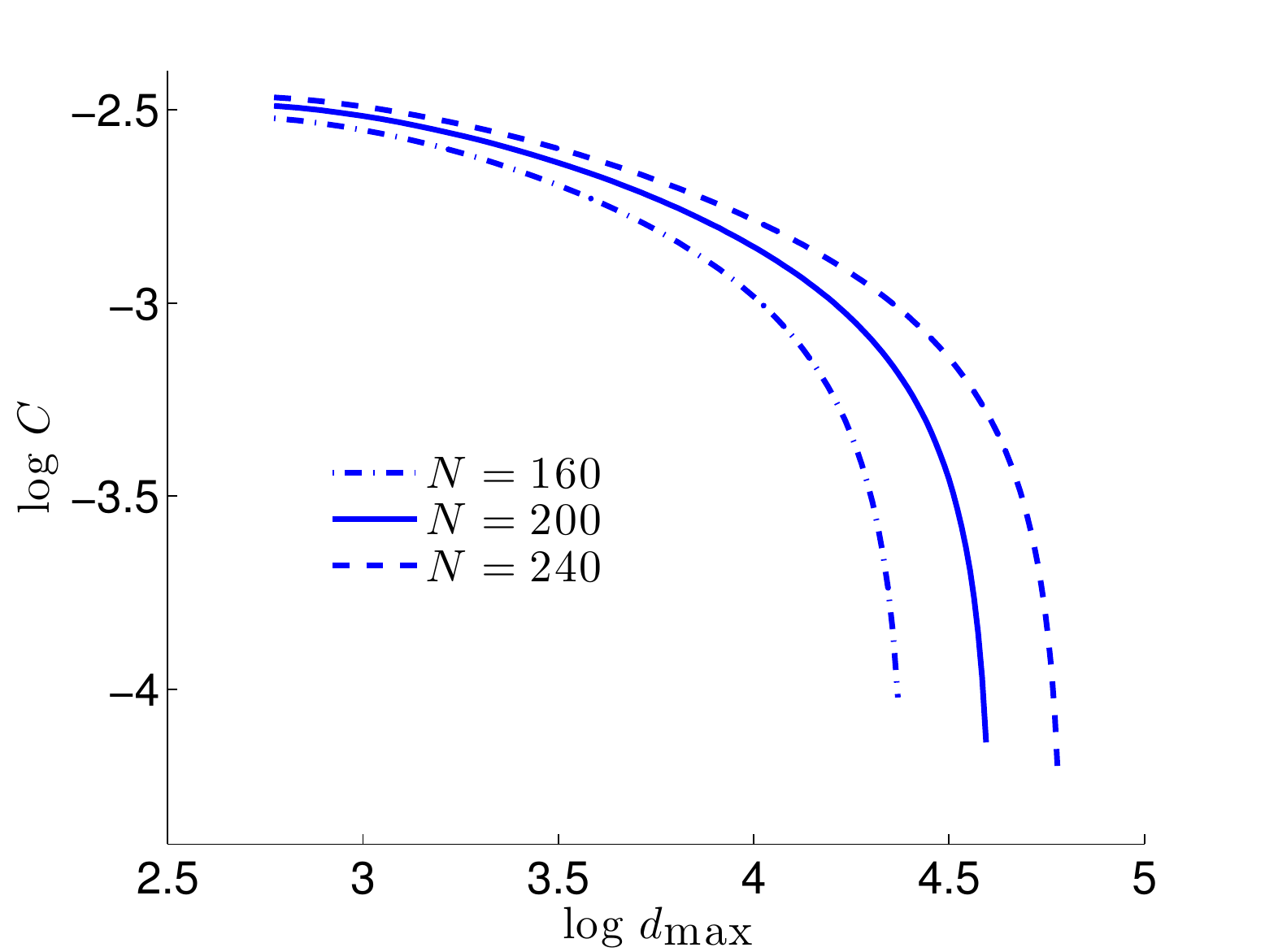}
\caption{\emph{Correlations in the out-of-equilibrium regime.} Both panels show the correlations in the time window $d_\min < vt \ll d_\max$, as a function of $d_\max$. The upper panel considers the case in which both points lie in the same half, showing a power-law decay with $d_\max$. The lower panel considers the case in which the two points lie in different halves, showing no power-law behavior. In both cases, we have considered $d_\min=1$, $16<d_\max<N/2$ and $t=3\;(1/J)$.} 
\label{plot_corre_small_times}
\end{figure}

Let $d_1$ and $d_2$ be the distances from both points to the $LR$ interface, $d_\min=\min(d_1,d_2)$ and $d_\max=\max(d_1,d_2)$. The out-of-equilibrium regime is defined by the condition $d_\min < vt <d_\max$, i.e.: the time lapse in which the quasi-particles have already reached the closest point and have not yet left the region between the two points. In the time-regime where $d_\min<vt\ll d_\max$, the CFT predicts that after a joining quench the correlation function will behave as $C \propto d_\max^{-2x-1/2}$ \cite{c&c2}, independently on whether the points are in the same or different halves. This prediction has been confirmed numerically \cite{divakaran2011non}. Fig. \ref{plot_corre_small_times} shows the results in our case. After the split quench, when the points are in the same half, we also observe $C \propto d_\max^{-\alpha}$, but with $\alpha\approx 1/2$ ($\alpha=0.46(5)$). Generalizing the CFT prediction, we may write this as $C \propto d_\max^{-2x-1/4}$. In the case of points in different halves, we do not observe any power law decay in the correlation function, as shown in the lower panel of Fig. \ref{plot_corre_small_times}. 

\begin{figure}[h!]
\includegraphics[width=\columnwidth]{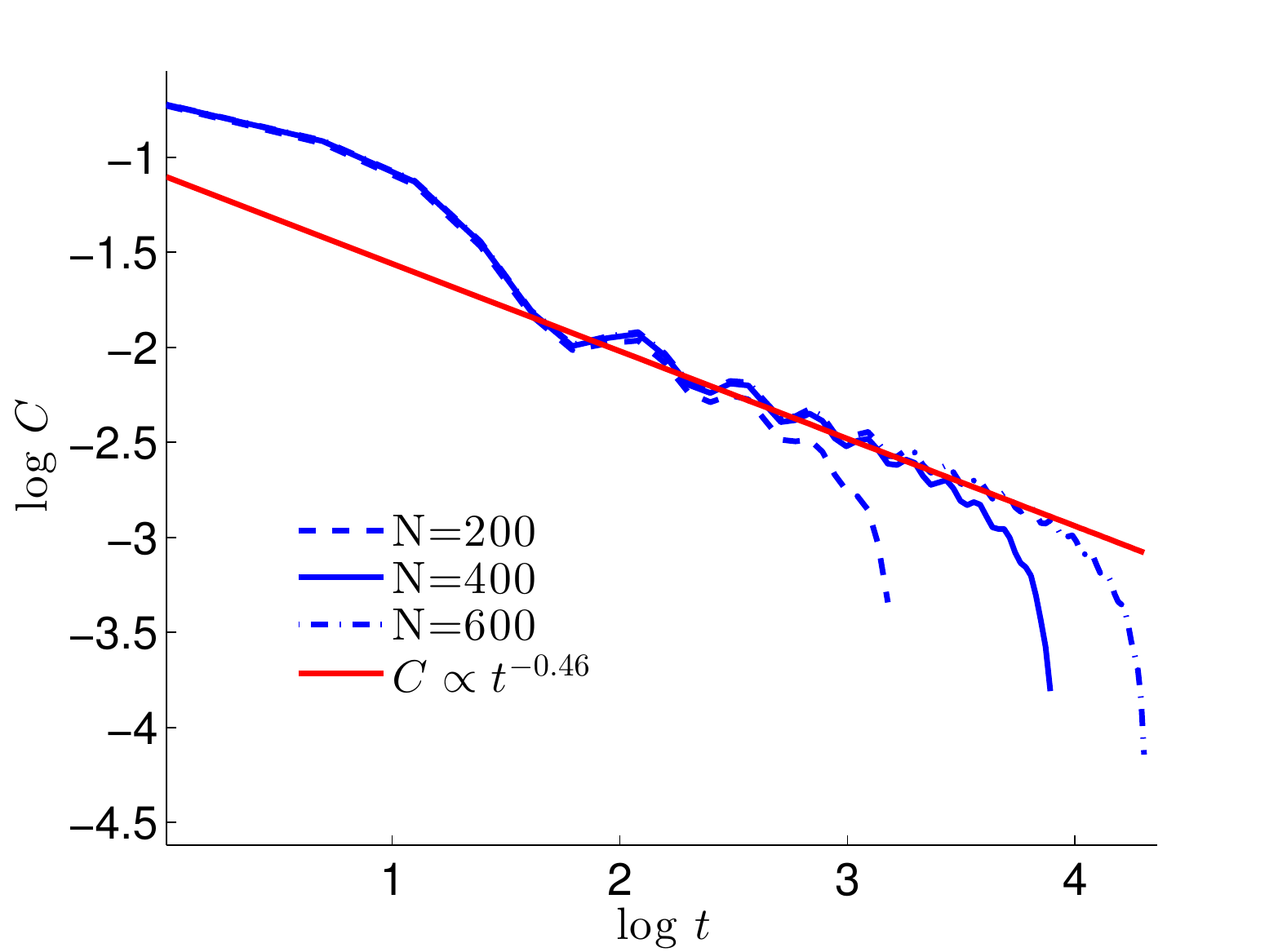}
\includegraphics[width=\columnwidth]{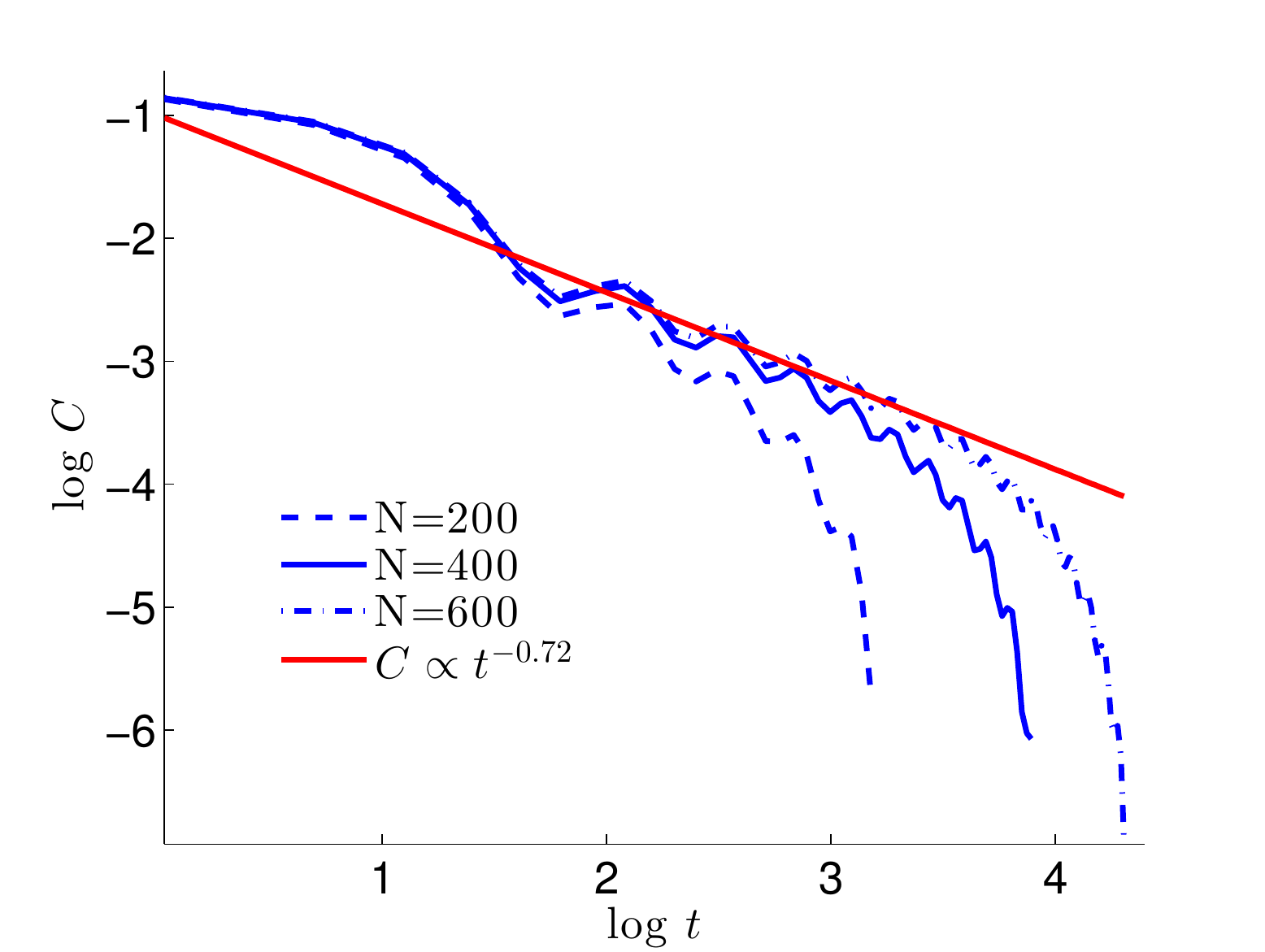}
\caption{\emph{Correlations in the light-cone regime.} We measure the correlations between two points in the regime described in the text as \emph{light-cone}, with $\frac{d_\min}{vt}=1/2$, $\frac{d_\max}{vt}=2$ and $vt<N/2$. Upper panel: if both points are in the same half, correlations decay with time as $t^{-3x}$ with $x=0.46(1)$. Lower panel: if both points are in different halves, correlations decay with the same law, but  $x=0.72(1)$, close to $3/4$.}
\label{plot_corre_log_times}
\end{figure}

We study next a kind of \emph{light-cone regime} by fixing the ratios $\epsilon(t)=\frac{d_\min}{vt}<1$ and $R(t)=\frac{d_\max}{vt}>1$. In the joining quench, the correlation in this regime is described by $C \propto t^{-3x}$ as $t\to\infty$ \cite{divakaran2011non}. Our results for the split quench are shown in Fig. \ref{plot_corre_log_times}. We can observe also a power law decay of correlations as in the joining case but when both points are located in the same half of the chain the exponent we extract is $0.46(1)$ that apparently is not compatible with the joining case. Interestingly, when they lay in different halves, we still observe a polynomial decay of the correlations with $t$, but with the exponent close  $3/4$, $0.72(1)$, that doubles the one observed in the joining case.

Finally, we reach the equilibrium regime. If both points lie on the same half, correlations decay as a power law of the system size, as in the joining quench as shown in the inset of the upper panel of Fig. \ref{plot_corre_same_half}. This means that the system is still critical, with the same critical exponent $x=1/8$, and has thermalized to a temperature which is very close to zero. If the points are on different halves we still observe  power-law behavior but  this time we observe anti-correlations that  decay with the same critical exponent $x=1/8$,  as shown the inset in the lower panel of Fig. \ref{plot_corre_same_half}. 

\subsection{Local properties}
\label{sec_local_properties}

As we have seen, the evolution of global quantities after the split quench is very different  from the one after a  join quench. This is not very surprising, since both quenches are globally very different. Still, we can attempt a local characterization of the equilibrium regime. Since the quenching Hamiltonian are locally identical, and the initial states provide the same correlations functions in the bulk,  one might expect similar behaviors in both quenches. 

\begin{figure}[h!]
\centering
\subfigure{\includegraphics[width=\columnwidth]{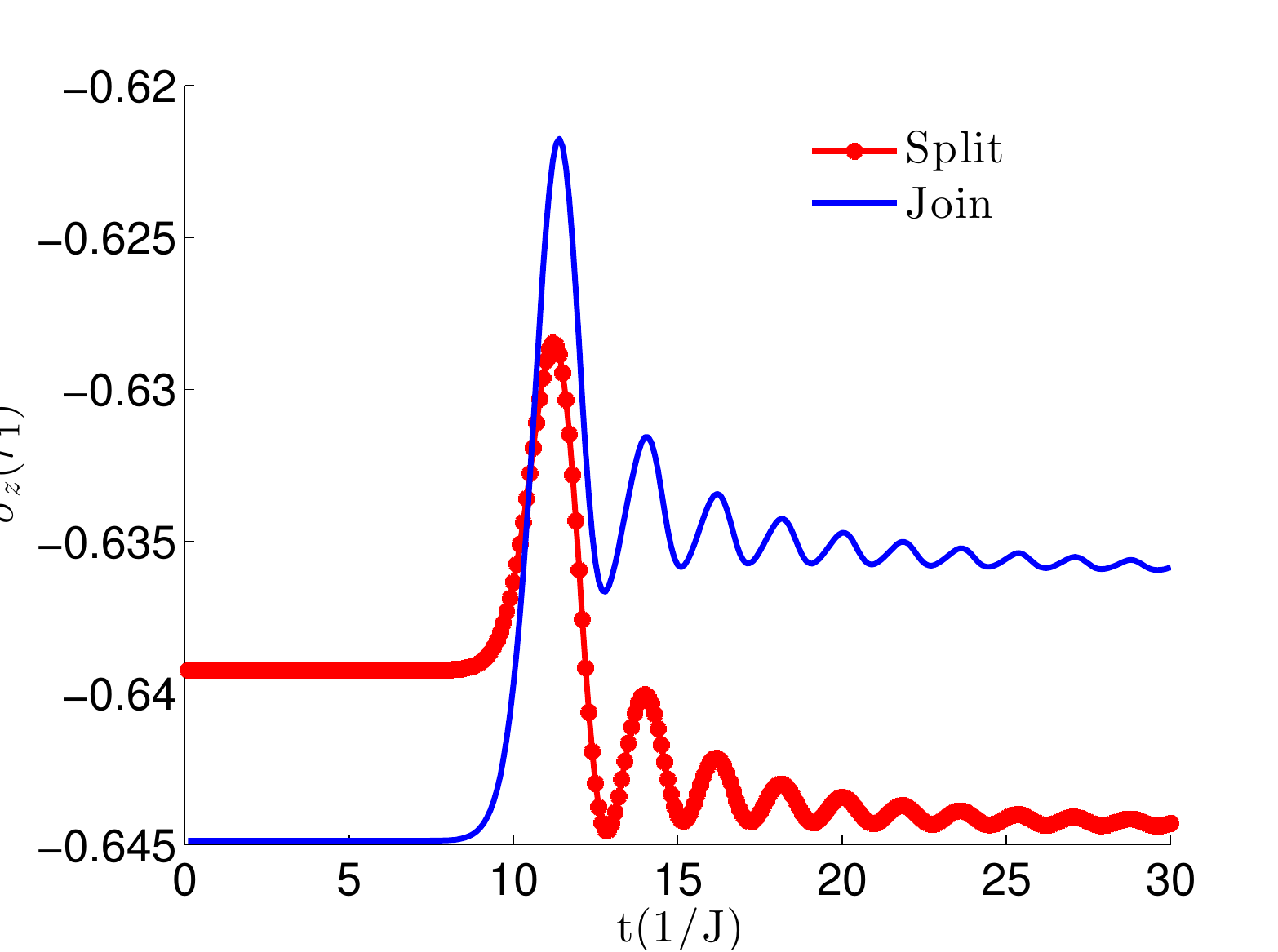}}
\subfigure{\includegraphics[width=\columnwidth]{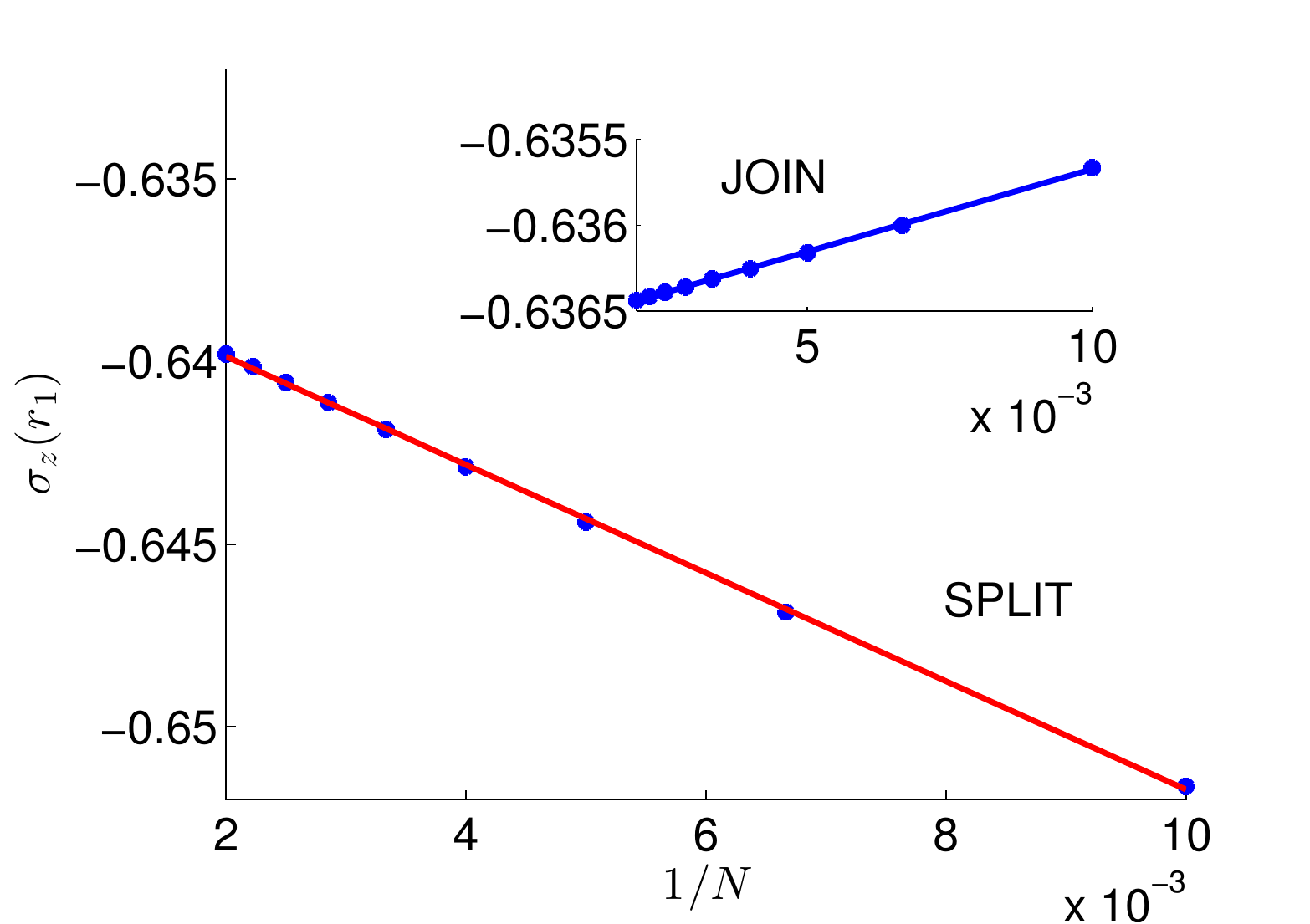}}
\caption{\emph{Top:} Time-evolution of the expected magnetization at a site $r=N/2-N/10$, both after a split quench of an initial chain with $N=200$, and after a join quench of two initial chains with size $100$. During the out-of-equilibrium regime, the join and the split values interchange and finally relax to different equilibrium values. \emph{Bottom:} Scaling analysis shows that in both cases the equilibrium value is well described by Eq. \ref{eq:magx}, but the thermodynamic limit is approached from different directions, $\tilde{c}_z^{\textrm{split}}<0$, while $\tilde{c}_z^{\textrm{join}}>0$, implying that even at a local level the two quench protocols are well distinguishable.\label{plot_mag}}
\end{figure}

The top panel of Fig. \ref{plot_mag} shows that, as expected, after splitting or joining,  local observables display all three stages, the static, the out-of-equilibrium and the equilibrium stages.
The static value of the magnetization  depends on the system size as 

\begin{equation}
\langle \sigma_z \rangle_0=\sigma_\infty+\frac{c_z}{N}, 
\label{eq:magx}
\end{equation}
where both $\sigma_\infty$ and $c_z$ are known analytically \cite{burkhardt_finite-size_1985}. Let us consider the two  cases of a $N=200$ spin-chain split into two halves  and two $N=100$ chains joined in a single chain. In both cases the static values differ because of the different initial system sizes, with the split value displaying larger magnetization than the join. As expected, both values cross during the out-of-equilibrium phase, and the split equilibrium magnetization  is lower than the join equilibrium magnetization. The equilibrium magnetization for the split chain converges to the static value of the two chains that have been joined, while the opposite does not happen, the equilibrium value for the joint chain is not the same as the static value of larger chain before the split. This is a finite-size effect. In the thermodynamic limit, the magnetization is the same before and after the quench in both cases. Still, for finite chains, we can distinguish both quench protocols, since the magnetization approaches the thermodynamic limit from opposite directions.

\begin{figure}[h!]
\centering
\subfigure{\includegraphics[width=\columnwidth]{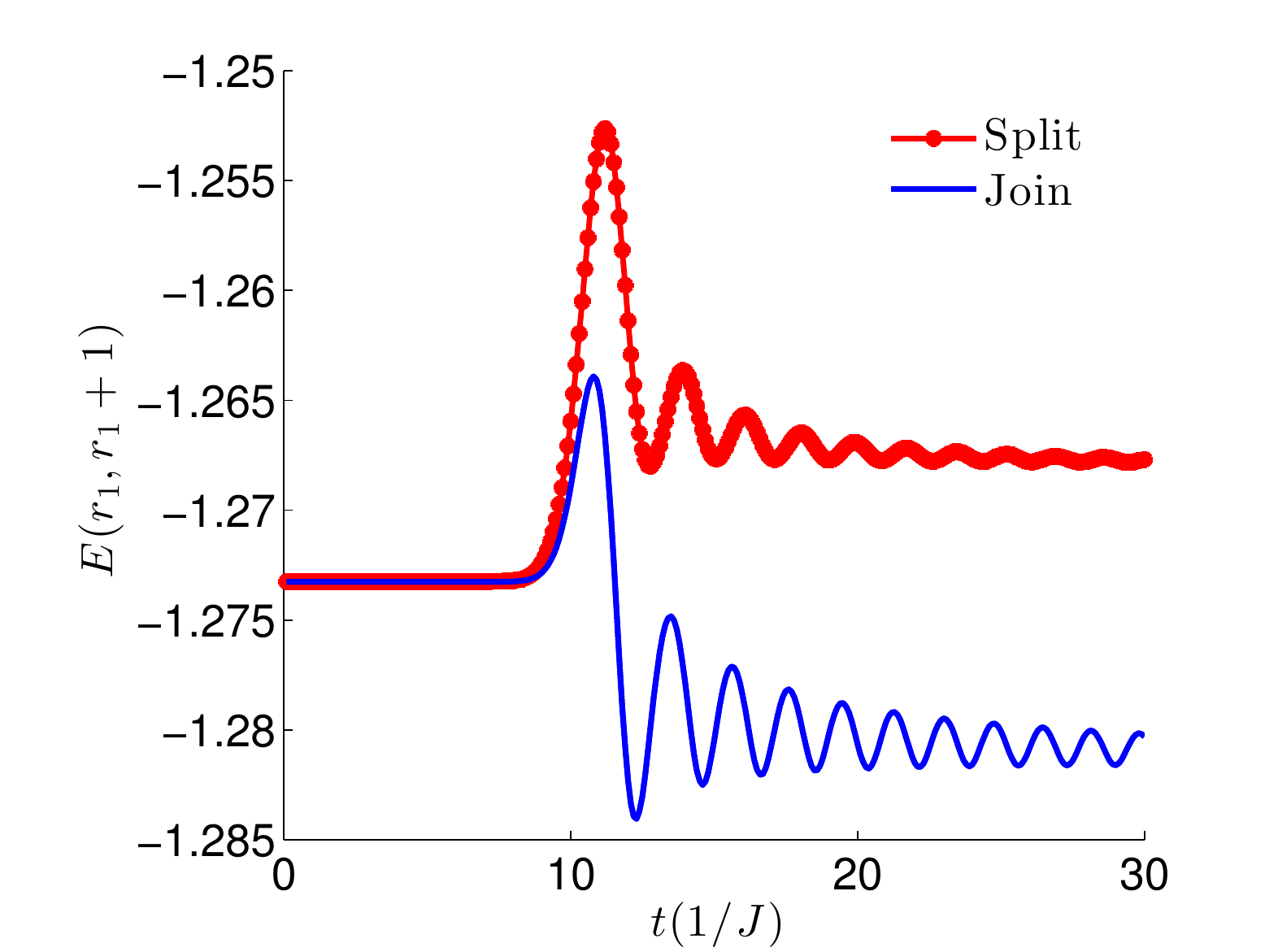}}
\subfigure{\includegraphics[width=\columnwidth]{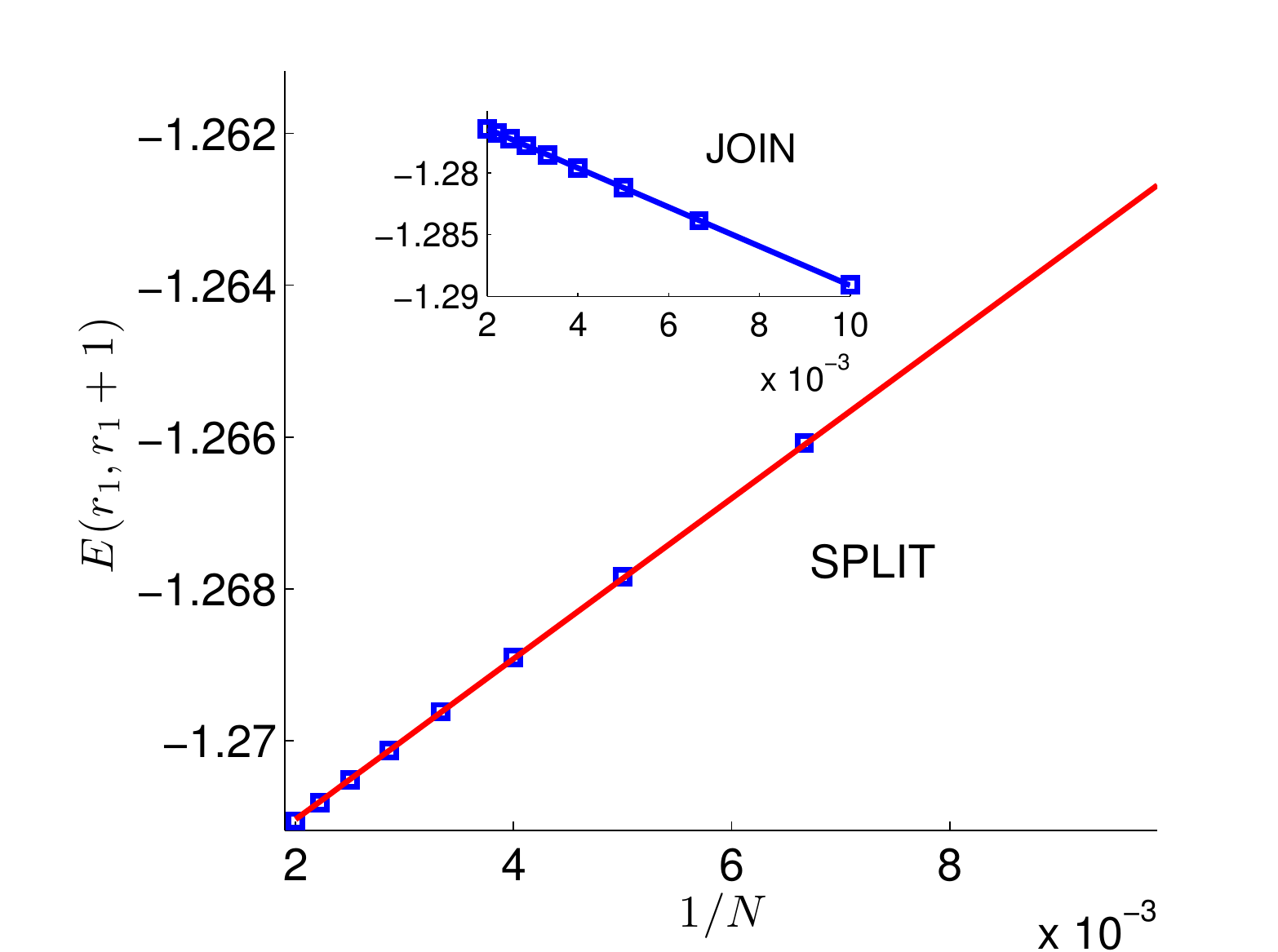}}
\caption{\emph{Top:} Time-evolution of the expected energy at link $r_1=N/2-N/10$, both after a split quench of an initial chain with $N=200$, and after a join quench of two initial chains with size $100$. They relax to different equilibrium values. \emph{Bottom:} The scaling analysis shows that in both cases the equilibrium value is the same but the thermodynamic limit is approached from different directions, implying that even at a local level the two quench protocols are well distinguishable.\label{plot_ene}}
\end{figure}

Indeed, as shown in the bottom panel of Fig. \ref{plot_mag}, the fit to Eq. \ref{eq:magx} of the finite-size data shows that $\tilde{c}_z^{\textrm{split}}< 0$ and $\tilde{c}_z^{\textrm{join}} >0$, implying that even at a local level the two quench protocols are well distinguishable.
{The same study is performed for the energy density in Fig. \ref{plot_ene}, where again we see that the quenches are completely distinguishable at a local level.}

\begin{figure}[h!]
\includegraphics[width=\columnwidth]{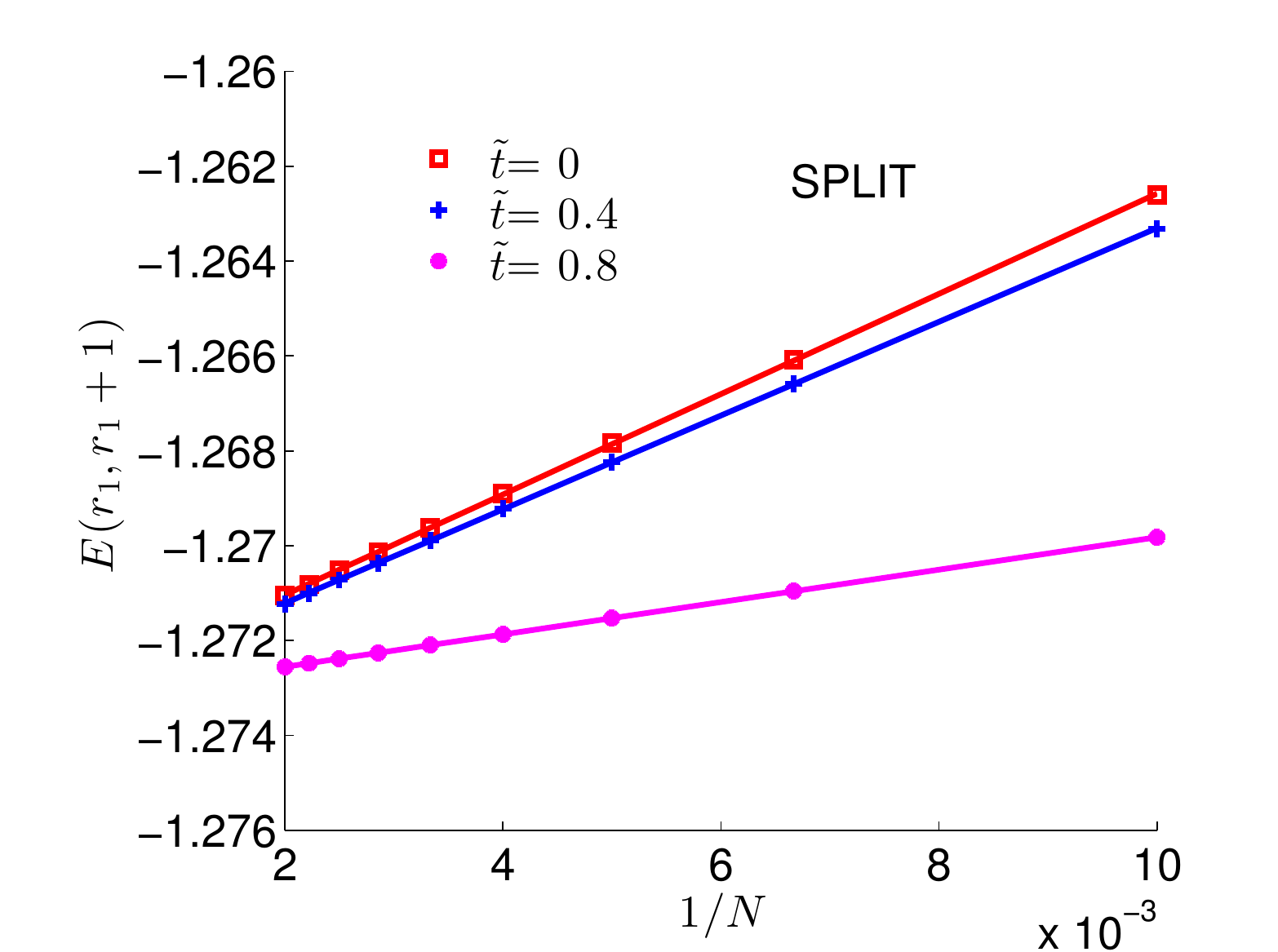}
\includegraphics[width=\columnwidth]{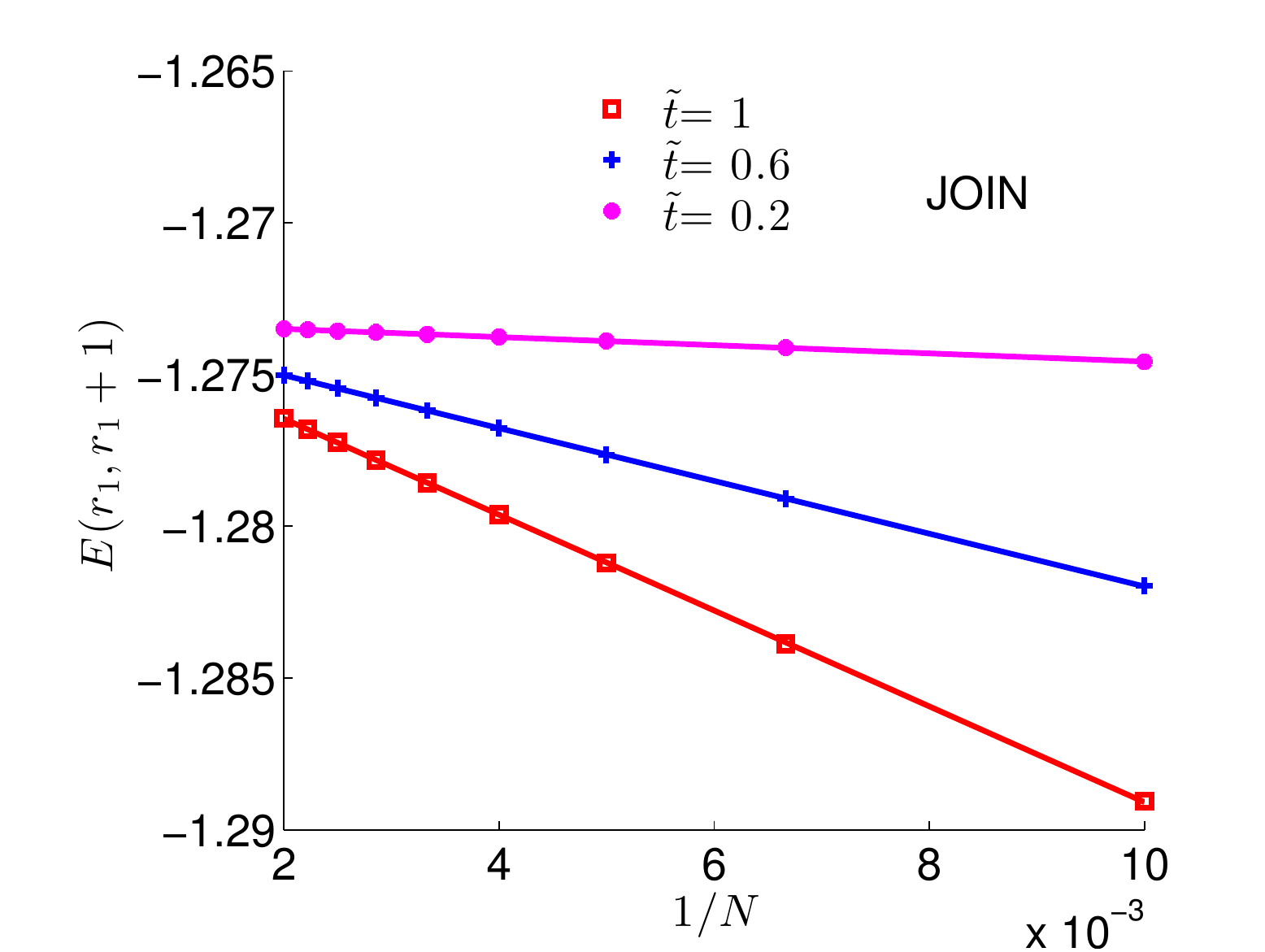}
 \caption{Intermediate quenches. The scaling of local observable with respect to the system size are presented in the equilibrium regime after either weakening the Hamiltonian between $L$ and $R$ so to provide a partial split of the two originally joined chains  {(\emph{upper panel})} and after introducing a weak bond between $L$ and $R$ so to partially join the originally separated  chains {(\emph{lower panel})}. The slope of the finite size effects is in one to one correspondence with the strength of the Hamiltonian bond joining $L$ and $R$ giving the possibility to local discern all the above scenarios. We have considered $r_1=N/2-N/10$. }
\label{int_quenches_ene}
\end{figure}

Finally we can also characterize intermediate quenches, considering a parameter $\tilde{t}$ that modifies the strength of the bond connecting $L$ and $R$, as $\tilde{t}\cdot H_{LR}$(see Eq. (\ref{eq:h0})) so that the quench is obtained by varying the initial value of  $ \tilde{t}$. In this way we can either weaken the Hamiltonian bond between $L$ and $R$ by passing from the initial value of $\tilde{t}=1$ to a quench value of  $0<\tilde{t}<1$ so to  partially split the chain. Alternatively we can quench from the initial $\tilde{t}=0$ to again any value $0<\tilde{t}<1$  so to partially join $L$ and $R$  by switching on  a weaker bond between them (weaker than the other present in the chain). 
The numerical results for such intermediate quenches are shown in Fig. \ref{int_quenches_ene} where we appreciate that by looking at the sign of the finite size corrections we can distinguish if the chain is being split (even partially) or joined. This last situation is similar to the study of the effects of impurities in critical systems \cite{eisler_2014}.

\section{Conclusions and outlook}
\label{sec:conclusions}
In this work we have discussed the time evolution of a critical spin-chain which is quenched by effectively disconnecting its two halves. Due to the entanglement in the initial state each of the two halves is originally in a mixed state. The bipartite entanglement between the halves is conserved during the evolution and so is the entanglement spectrum of the bipartition. 
We address the role of the conservation of the ES by comparing this quench with the one where two independent spin chains are joined together. Both the initial state and the quenching Hamiltonian are locally  indistinguishable in the bulk of the system away from the partition in two halves. The joining quench however, due to the interaction between the two halves, does not conserve the entanglement spectrum of the bipartition. We show that the equilibrium states emerging after the two quenches differ both globally and locally.

This suggests that the conservation of the entanglement spectrum has important consequences on both the out-of-equilibrium evolution of many-body systems and their equilibration regime.

As opposed to other scenarios discussed in the literature, the conservation of the ES is not related to integrability of the dynamics, but rather to the specific quench protocol  and the basic nature of entanglement. 

The splitting  of a spin chain in two halves is a local quench  and as such does not inject enough energy in the system to observe thermalization at any non-zero temperature. In other terms, it can not give rise to an equilibrium state with finite entropy density.

We plan to generalize this analysis to global quenches that inject enough energy in the initial state  for effective thermalization to take place an thus address which are the effects of the conservation of the  entanglement spectrum also in those scenarios. 

\section{Acknowledgments}

We would like to thank P. Calabrese, M. Fagotti, E. Tonni for numerous discussions on the topic. In particular we thank  M. Fagotti  and A. Riera for their critical reading of a first version of our manuscript. We acknowledge  financial support from the Marie Curie project FP7-PEOPLE-2010-IIF ENGAGES 273524, TOQATA  (FIS2008-00784), ERC QUAGATUA OSYRIS,  EU IP SIQS, and the Spanish government grant FIS2012-33642.




\newpage
\appendix

\section{Profile of the entanglement entropy}
\label{app_entropy_scaling}

We guess from our numerical data  the following functional form  for the projection of the entanglement entropy $S(r,t)$ over space or time discussed in Sect. IIIB:

\begin{align}
&\tilde{S}_X(x)=\frac{c}{\alpha_x}\log_2\left|\frac{N}{\pi} \left( \sin\frac{2\pi x}{T_x} \right)^{\nu_x} \right|+\textrm{cst}.
\label{eq_ent_projection_appendix}
\end{align}
In this way by either chosing $x=r$ or $x=t$ we recover both space and time sections of the entanglement entropy that are presented in Fig. \ref{plot_entr_projection}. In each case $T_x$ has to be chosen accordingly to the definition discussed in the next paragraph.
\begin{align}
&x=r \to \tilde{S}_X(x)=\tilde{S}_S(r) \nonumber \\
&x=t \to \tilde{S}_X(x)=\tilde{S}_T(t) 
\end{align}
\\

\begin{itemize}
\item \textbf{Calculation of the parameter $\alpha_x$.}
We first determine the value of $\alpha_x$  in both cases. 
 $\alpha_x$ is extracted through a finite size scaling analysis. By  fxing the ratio $\frac{t}{N}$, we study how  Eq. (\ref{eq_ent_projection_appendix}) depends on $N$ since  $\tilde{S}_X(x)=\frac{c}{\alpha_x}\log_2 N+\textrm{cst}$. Specifically the calculation has been done using chains with  $N=100,120,140,160,200,240$.
\\

For the case of $\tilde{S}_T(t)$ we have considered ratios  
\begin{equation}
\frac{t}{N}=\frac{1}{2}\cdot\left( \frac{2}{10},\frac{3}{10},\frac{5}{10},\frac{6}{10} \right),
\end{equation}
while for $\tilde{S}_S(r)$ we have considered sites $r\ll N/2$ and ratios
\begin{equation}
\frac{t}{N}=\left( \frac{2}{10},\frac{3}{10},\frac{4}{10},\frac{6}{10},\frac{7}{10} \right).
\end{equation}
Averaging on the values of  $\alpha$ extracted from the various ratios we obtain 
\begin{align}
& \tilde{S}_T(t) \to \alpha_T=2.919\;(41) \nonumber \\
& \tilde{S}_S(r) \to \alpha_S=2.947\;(13) 
\end{align}
that are both compatible with $\alpha=3$. 
In order to cross-check our strategy we have repeated the same procedure for the join quench. 
In this case we obtain 
\begin{align}
& \tilde{S}_T(t) \to \alpha_{join}=3.012 \;(59) \nonumber \\
& \tilde{S}_S(r) \to \alpha_{join}=3.042 \;(10)
\label{eq_values_alpha_join} 
\end{align}
that is in agreement with the available theoretical prediction  $\alpha_{join}=3 $   \cite{calabrese2007entanglement,dubail,igloi2012entanglement}.

\item \textbf{Calculation of the parameter $\nu_x$.}

At fixed $N$, $\tilde{S}_X(x)$ depends linearly on $y=\log_2\left|\left( \sin\frac{2\pi x}{T_x} \right) \right| $. We can thus extract the value of $\nu_x$ throug a linear fit of $\tilde{S}_X$ as a function of $y$ at fixed $N$.
We have performed such analysis for several  chains of length $N=100,120,140,160,200,240$ obtaining several estimates of $\nu_x$.
\\

In particular for  $\tilde{S}_T(t)$, $T_x=N$, and for each $N$ we have considered the  set of times
\begin{equation}
t=\left( \frac{N}{8}, \frac{N}{8}+1,...,\frac{N}{2}-\frac{N}{8}  \right).
\end{equation}
For $\tilde{S}_S(r)$, $T_x=2N$ and we have considered $r\ll N/2$:
\begin{equation}
r=\left( \frac{2N}{10}, \frac{2N}{10}+1,...,\frac{N}{4}  \right)
\end{equation}

Averaging on the values of  $\nu$ obtained for each $N$ we get
\begin{align}
& \tilde{S}_T(t) \to \nu_T=0.500 \; (14) \label{res:str}\\
& \tilde{S}_S(r) \to \nu_S=1.004 \; (5) 
\label{eq_values_alpha_split} 
\end{align}
While the first result is un-expected, the second is compatible with what expected in a join quench.  By repeating the analysis for the join quench where we expect  $\nu_{join}=1$ \cite{calabrese2007entanglement,dubail,igloi2012entanglement} we indeed find 
\begin{align}
& \tilde{S}_T (t) \to \nu_{join}=0.975 \; (34) \nonumber \\
& \tilde{S}_S (r) \to \nu_{join}=1.070 \; (10)
\end{align}
as expected.
\end{itemize}


\begin{thebibliography}{10}

\expandafter\ifx\csname url\endcsname\relax
  \def\url#1{{\tt #1}}\fi
\expandafter\ifx\csname urlprefix\endcsname\relax\def\urlprefix{URL }\fi
\providecommand{\eprint}[2][]{\url{#2}}


\bibitem{von_neumann.29} J. von Neumann, Z. Phys {\bf 57}, 30 (1929).

\bibitem{lewenstein.book} M.~Lewenstein, A.~Sanpera and V.~Ahufinger,
  {\em Ultracold atoms in optical lattices}, Oxford University Press
  (2012).
\bibitem{kinoshita_quantum_2006} T. Kinoshita, T. Wenger and D.S. Weiss,
  Nature {\bf 440}, 900 (2006).

\bibitem{cheneau_light-cone-like_2012} M. Cheneau, P. Barmettler,
  D. Poletti, M. Endres, P. Schausz, T. Fukuhara, C. Gross, I. Bloch,
  C. Kollath and S. Kuhr, Nature {\bf 481}, 484 (2012).

\bibitem{gring_relaxation_2012} M. Gring, M. Kuhnert, T. Langen,
  T. Kitagawa, B. Rauer, M. Schreitl, I. Mazets, D.A. Smith, E. Demler
  and J. Schmiedmayer, Science {\bf 337}, 1318 (2012).

\bibitem{trotzky_probing_2012} S. Trotzky, Y.A. Chen, A. Flesch,
  I.P. McCulloch, U. Schollw\"ock, J.  Eisert and I. Bloch,
  Nat. Phys. {\bf 8}, 325(2012).

\bibitem{langen_local_2013} T. Langen, R. Geiger, M. Kuhnert, B. Rauer
  and J. Schmiedmayer, Nat. Phys. {\bf 9}, 640 (2013)
\bibitem{deutsch.91} J.M. Deutsch, Phys. Rev. A {\bf 43}, 2046 (1991).

\bibitem{polkovnikov.11} A. Polkovnikov, K. Sengupta, A. Silva and
  M. Vengalattore, Rev. Mod. Phys. {\bf 83}, 863 (2011).

\bibitem{sredniki.94} M. Sredniki, Phys. Rev. E {\bf 50}, 888 (1994).

\bibitem{rigol.08} M. Rigol, V. Dunjko, and M. Olshanii, Nature {\bf
  452}, 854 (2008); M. Rigol, Phys. Rev. Lett. {\bf 103}, 100403
  (2009). 

\bibitem{jaynes.57} E. Jaynes, Phys. Rev. {\bf 106}, 620 (1957).

\bibitem{landau_statistical_1980} L.D. Landau, L.P. Pitaevskii and
  E.M. Lifshitz, {\em Statistical physics I}, Pergamon Press (1980).

\bibitem{pathria_statistical_1996} R.K. Pathria, {\em Statistical
  mechanics}, Butterworth-Heinemann (1996).

\bibitem{balian_microphysics_2007} R. Balian, {\em From Microphysics
  to Macrophysics: Methods and Applications of Statistical Physics},
  Springer (2007).

\bibitem{rigol2007relaxation} M. Rigol, V. Dunjko, V. Yurovsky and
  M. Olshanii, Phys. Rev. Lett. {\bf 98}, 050405 (2007).



\bibitem{berges2004prethermalization} J. Berges, S. Borsanyi and
  C. Wetterich, Phys. Rev. Lett. {\bf 93}, 142002 (2004).

\bibitem{kollar_generalized_2011} M. Kollar, F.A. Wolf and
  M. Eckstein, Phys. Rev. B {\bf 84}, 054304 (2011).

\bibitem{li_entanglement_2008} H. Li and F.D.M. Haldane,
  Phys. Rev. Lett. {\bf 101}, 010504 (2008).
\bibitem{cazalilla} M. Chung, A. Iucci, M. A. Cazalilla, New J. Phys. {\bf 14}  075013 (2012).
M. A. Cazalilla, A. Iucci, M. Chung, Phys. Rev. E {\bf 85}, 011133 (2012).
  
\bibitem{popescu_entanglement_2006} S. Popescu, A.J. Short and
  A. Winter, Nat. Phys. {\bf 2}, 754 (2006).

\bibitem{riera_thermalization_2012}
A. Riera, C. Gogolin and J. Eisert, Phys. Rev. Lett. {\bf 108}, 080402 (2012).

\bibitem{masanes_complexity_2013}
L. Masanes, A.J. Roncaglia and A. Ac\'{\i}n, Phys. Rev. E {\bf 87}, 032137 (2013).

\bibitem{fagotti_reduced_2013} M. Fagotti and F.H.L. Essler, Phys. Rev. B {\bf 87}, 245107 (2013).

\bibitem{hastings_area_2007} M.B. Hastings, J. Stat. Mech.:
  Theor. Exp. P08024 (2007).

\bibitem{masanes_area_2009} L. Masanes, Phys. Rev. A {\bf 80}, 052104
  (2009).

\bibitem{calabrese2005evolution} P. Calabrese and J. Cardy,
  J. Stat. Mech.: Theor. Exp.  P04010 (2005).

\bibitem{hauke_spread_2013} P. Hauke and L. Tagliacozzo,
  Phys. Rev. Lett. {\bf 111}, 207202 (2013).

\bibitem{gogolin_11} C. Gogolin, M.P. M\"uller and J. Eisert, Phys. Rev. Lett. {\bf 106}, 140401 (2011).

\bibitem{schachenmayer_entanglement_2013} J. Schachenmayer,
  B.P. Lanyon, C.F. Roos and A.J. Daley, Phys. Rev. X {\bf 3}, 031015
  (2013).

\bibitem{c&c2} P. Calabrese and J. Cardy, Phys. Rev. Lett. {\bf
  96}, 136801 (2006).

\bibitem{dubail} J.M. St{\'e}phan and J. Dubail, J. Stat. Mech.:
  Theor. Exp. P08019 (2011).

\bibitem{peschel07} V. Eisler and I. Peschel, J. Stat. Mech.: Theor. Exp. P06005
  (2007).

\bibitem{lieb1964two} E. Lieb, T. Schultz and D. Mattis, Rev. Mod. Phys {\bf 36}, 856 (1964).

\bibitem{fagotti2008evolution} M. Fagotti and P. Calabrese, Phys. Rev. A {\bf 78}, 010306 (2008)

\bibitem{torlai_dynamics_2013} G. Torlai, L. Tagliacozzo and G. De
  Chiara, J. Stat. Mech.: Theor. Exp. P06001 (2014).

\bibitem{vidal_efficient_2003} G. Vidal, Phys. Rev. Lett. {\bf 91}, 147902 (2003). 

\bibitem{vidal_efficient_2004} G. Vidal, Phys. Rev. Lett. {\bf 93}, 040502 (2004).

\bibitem{cirac_entanglement_2011} J.I. Cirac, D. Poilblanc, N. Schuch
  and F. Verstraete, Phys. Rev. B {\bf 83}, 245134 (2011).

\bibitem{peschel_corner_1987} I. Peschel and T.T. Truong, Z. Phys. B {\bf 69}, 385 (1987). 

\bibitem{lauchli_operator_2013} A.M. L\"auchli, ArXiv: 1303.0741 (2013).

\bibitem{de_chiara_entanglement_2012} G. De Chiara, L. Lepori,
  M. Lewenstein and A. Sanpera, Phys. Rev. Lett. {\bf 109},
  237208 (2012).
\bibitem{alba_2012} V. Alba, M. Haque and A. L\"auchli, Phys. Rev. Lett. {\bf 108}, 227201 (2012)

\bibitem{holzhey1994geometric} C. Holzhey, F. Larsen and F. Wilczek,
  Nucl. Phys. B {\bf 424}, 443 (1994).

\bibitem{callan1994geometric} C. Callan and F. Wilczek, Phys. Lett. B {\bf 333}, 55 (1994).

\bibitem{srednicki1993entropy} M. Srednicki, Phys. Rev. Lett. {\bf 71}, 666 (1993).

\bibitem{vidal_latorre_rico_kitaev} G. Vidal, J.I. Latorre, E. Rico
  and A. Kitaev, Phys. Rev. Lett. {\bf 90}, 227902 (2003).

\bibitem{c&c} P. Calabrese and J. Cardy, J. Stat. Mech.: Theor. Exp. P06002 (2004).

\bibitem{calabrese2007entanglement} P. Calabrese and J. Cardy, J. Stat. Mech.: Theor. Exp. P10004 (2007)

\bibitem{igloi2012entanglement} F. Igl{\'o}i, Z. Szatm{\'a}ri and
  Y.C. Lin, Phys. Rev. B {\bf 85}, 094417 (2012).

\bibitem{eisler2008entanglement} V. Eisler, D. Karevski, T. Platini and
I. Peschel, J. Stat. Mech.: Theor. Exp. P01023 (2008).

\bibitem{lauchli2008spreading} A.M. L{\"a}uchli and C. Kollath, J. Stat. Mech.: Theor. Exp. P05018 (2008).

\bibitem{collura2013entanglement} M. Collura and P. Calabrese, J. of Phys. A: Math. Theor. {\bf 46}, 175001 (2013).
\bibitem{eisler_2014} V. Eisler and I. Peschel J. Stat. Mech. : Theor. Exp. P04005 (2014).
\bibitem{alba_2014}  V.Alba and F. Heidrich-Meisner arXiv:1402.2299.
\bibitem{perales2008entanglement} \'A. Perales and G. Vidal, Phys. Rev. A {\bf 78}, 042337 (2008).

\bibitem{divakaran2011non} U. Divakaran, F. Igl{\'o}i and H. Rieger,
J. Stat. Mech.: Theor. Exp. P10027 (2011).

\bibitem{burkhardt_finite-size_1985} T.W. Burkhardt and I. Guim, 
J. Phys. A: Math. Gen. {\bf 18}, L33 (1985).


\end{thebibliography}
\end{document}